\documentclass[12pt]{iopart}

\usepackage[hidelinks]{hyperref}
\usepackage{float}			
\usepackage{pgfplots}
\usepackage{caption}		
\usepackage{subcaption}
\usepackage{siunitx}
\usepackage{multirow}		

\begin{document}

\newcommand{\at}{\left. \vphantom{\frac{1}{1}} \right|}	

\title[Development and validation of an electron temperature-dependent interaction potential for silicon and copper for the use in atomistic simulations of laser ablation]{Development and validation of an electron temperature-dependent interaction potential for silicon and copper for the use in atomistic simulations of laser ablation}

\author{Simon Kümmel, Johannes Roth}

\address{Institute for Functional Matter and Quantum Technologies University of Stuttgart, Pfaffenwaldring 57, D-70569 Stuttgart}
\ead{simon.kuemmel@fmq.uni-stuttgart.de}
\vspace{10pt}
\begin{indented}
\item[]June 2025
\end{indented}

\begin{abstract}
Laser pulses with a duration of the order of femtoseconds lead to a strong excitation, heating and potentially to ablation of the irradiated material. During the time of strong excitation, the interaction of the atoms and thus the material dynamics can be strongly altered. To take this effect into account, an ip was developed for copper that takes the excitation of the electrons into account up to an electron temperature of \SI{1.2}{eV}. Furthermore, several ways to identify non-thermal effects in Density Functional Theory calculations and how to incorporate and validate them in molecular dynamics simulations are presented. Explicitly, the free energy curves, elastic constants and phonon spectra are compared. Additionally, it is shown that the change of the melting temperature with the degree of excitation is consistent with all of these properties. \\
Moreover, the behaviour of copper upon excitation is compared to silicon by using a similar potential that was previously developed by a different author.
\end{abstract}

\section{Introduction}
Density Functional Theory (DFT) calculations allow a very accurate description of matter with only few assumptions and approximations. However, studying the atomistic dynamics on time scales of picoseconds and length scales up to micrometers is unfeasible due to the comparably poor scaling of the method. Classical Molecular Dynamics (MD) simulations scale much better but require ips (IP) that are designed to reproduce the properties of the material under investigation. These IPs can be optimized using experimental data or data from DFT calculations. \\
Within Two-Temperature Model (TTM) simulations, the heat transfer from electrons to the lattice following laser excitation can be studied via coupled heat equations \cite{Anisimov_1974_TTM,Anisimov_1997_TTM}. This method is based on the assumption that the energies of the electronic system and the lattice can be described by two separate temperatures, the lattice temperature and the electron temperature. Now, in TTM-MD simulations, the motion of the atoms in MD simulations can be coupled to a differential equation that describes the heat conduction of the electrons \cite{Schaefer_2002_TTM,Ivanov_2003_TTM}. This enables the study of materials under strong laser irradiation, such as femtosecond laser pulses. \\
Both, from experiments \cite{van_Vechten_1979_nonthermal_Si,Medvedev_2015_nonthermal_Si,Ernstorfer_2009_FT_experiment} and DFT calculations \cite{Recoules_2006_FTDFT,Khakshouri_2008_FTMD_W,Murphy_2015_FTMD_W,Norman_2012_FTMD_Au,Migdal_2015_FTDFT,Daraszewicz_2023_FTMD_Au,Tanaka_2022_FTMD_Cu}, it is known that the potential energy landscape can change drastically upon excitation of the electrons due to laser irradiation. While neglecting this change of the interaction might be reasonable in certain materials or under certain conditions, it is unclear what consequences the change of the interaction has on the material dynamics and melting behaviour. In particular for silicon, it was found that only those models which take this effect into account are capable of accurately reproducing experimental ablation depth and efficiency \cite{Klein_2023_Dissertation}. Also, the electron pressure and subsequent blast force due to electron temperature gradients in the material play an important role in the dynamics of the material \cite{Falkovsky_1999_electron_pressure,Eisfeld_2018_electron_pressure,Eisfeld_2020_Dissertation,Winter_2017_electron_pressure}. \\
Up to now, electron temperature-dependent potentials for tungsten \cite{Khakshouri_2008_FTMD_W,Murphy_2015_FTMD_W}, gold \cite{Norman_2012_FTMD_Au,Daraszewicz_2023_FTMD_Au} and silicon \cite{Shokeen_2010_FTMD_Si,Kiselev_2016_HLRS,Kiselev_2017_Dissertation} as well as few other materials have been focused on in most previous investigations. Many of these IPs are limited in their usability since they were developed for one or several degrees of excitation with little or no possibility to interpolate between the potentials for a full dynamic calculation of the potential for the local degree of excitation at any given time. As far as the authors are aware, such simulations have only been conducted on silicon \cite{Kiselev_2016_HLRS,Kiselev_2017_Dissertation,Klein_2023_Dissertation,Bauerhenne_2024_TTM_MD}. In particular, the approach and potential developed by Kiselev et al. \cite{Kiselev_2016_HLRS, Kiselev_2017_Dissertation} allows an accurate description of arbitrary degrees of excitation in silicon. It is the only potential of which the authors are aware of that was used in large-scale atomistic simulations yielding excellent agreement of ablation depth and ablation efficiency with experiments \cite{Klein_2023_Dissertation}. This approach is therefore followed and applied to copper which is expected to have similar properties as the more frequently studied gold. A detailed description of the potential development process is presented and which properties can be compared to DFT data in order to assess the validity and usability of the potential. \\
The authors are aware of one electron temperature-dependent potential for copper \cite{Tanaka_2022_FTMD_Cu} using the same approach as in  \cite{Khakshouri_2008_FTMD_W}. While this potential can reproduce the change in bond strength, the potential development is by far not straightforward and is still limited in the accuracy of reproducing reference data such as the bulk modulus due to the assumptions made in the model. This, however, is crucial for an accurate description of the increasing pressure at an irradiated surface shortly after the laser irradiation. Here, a more straightforward and flexible approach is presented that can in principle be applied to a broader field of materials with higher accuracy. \\

This paper is organised as follows: First, the workflow of developing an electron temperature-dependent potential is introduced by explaining finite-temperature DFT, how the reference data was chosen and weighted and how this approach was applied to copper. Next, a comparison of ground state properties is provided that any interaction potentials should be able to reproduce. Finally, properties that change with increasing electron temperature are shown and discussed before concluding this work.

\section{Finite-temperature Density Functional Theory}

Laser irradiation of any material strongly changes the occupation of states. First, the electrons are in a non-equilibrium state before thermalising within several femtoseconds \cite{Weber_2019_BOE_Cu}. At this point, the electrons can be well described by a Fermi-Dirac distribution. Since effects following the laser irradiation on time scales ranging from several femtoseconds to hundreds of picoseconds are of interest, it is assumed that the Fermi-Dirac distribution is a reasonable approximation of the actual distribution at any given point of time. \\
For all DFT calculations, VASP \cite{Kresse_1993_VASP,Kresse_1996_VASP_1,Kresse_1996_VASP_2} was used with PAW potentials \cite{Kresse_1999_PAW,Bloechl_1994_PAW}. The cohesive free energy functional was optimized for each DFT calculation which conceptually was first introduced by Mermin \cite{Mermin_1965_FTDFT} for finite-temperature DFT calculations as an extension to the ground state DFT \cite{Hohenberg_1964_DFT,Kohn_1965_DFT}
\begin{equation}
  \eqalign{F \left[ n \left( \vec{r} \right) \right] =& T \left[ n \left( \vec{r} \right) \right] + \frac{1}{2} \int \int \text{d} \vec{r} \text{d} \vec{r'} \frac{n \left( \vec{r} \right) n \left( \vec{r}' \right)}{\left| r - r' \right|} \cr
  &+ \int \text{d} \vec{r} v \left( \vec{r} \right) n \left( \vec{r} \right) + E_{\text{XC}} \left[ n \left( \vec{r} \right) \right] - \sum_{n} \sigma S \left( f_{n} \right)}
\end{equation}
with the electronic entropy
\begin{equation}
  S(f_{n}) = - \left[ f_{n} \ln \left( f_{n} \right) + \left( 1 - f_{n} \right) \ln \left( 1 - f_{n} \right) \right]
\end{equation}
and the Fermi-Dirac distribution
\begin{equation}
  f_{n} = f \left( \frac{\epsilon_{n} - \mu}{\sigma} \right) = \frac{1}{\exp \left[ \left( \epsilon_{n} - \mu \right) / \sigma + 1 \right]}.
\end{equation}
The excitation also changes the occupation of the states via
\begin{equation}
  n \left( \vec{r} \right) = \sum_{n} f_{n} \left| \Psi_{i} \left( \vec{r} \right) \right|^2,
\end{equation}
where the electron temperature
\begin{equation}
  \sigma = k_{\text{B}} T_{\text{e}}
\end{equation}
specifies the degree of excitation.

\section{Potential development}

\subsection{Overview}

For this work, the approach for the potential development by Kiselev et al. \cite{Kiselev_2017_Dissertation,Kiselev_2016_HLRS} was followed. In these works, a modified Tersoff potential \cite{Kumagai_2007_Tersoff_mod} for silicon was developed by fitting the cohesive free energy using potfit \cite{Brommer_2007_potfit}. potfit employs the force matching method which goes back to Ercolessi et al. \cite{Ercolessi_1994_Force_matching} and uses both simulated annealing \cite{Kirkpatrick_1983_Simulated_annealing} and a conjugate gradient algorithm using Powell's method \cite{Powell_1965_Powell_minimisation} for the optimisation of the potentials. This is done in order to find the best agreement with the reference data quickly and reliably. \\
As was done by Kiselev, first, independent potentials at various electron temperatures were developed. Next, all parameters are fitted to a polynomial of degree $N$
\begin{equation}
  P_{k} \left( T_{e} \right) = \sum_{n=0}^{N} p_{k,n} \left( \text{k}_{\text{B}} T_{e} \right)^{n}.
\end{equation}
Here, $p_{k, n}$ is the $n$$^{\text{th}}$ fitting parameter of $P_{k}$. For this to work, the parameters at various electron temperatures have to show a clear trend. Kiselev used the same approach for the four parameters $A$, $B$, $\lambda_{2}$ and $c_{2}$ while fixing the others to the unexcited values of the modified Tersoff potential. For the copper potential described in the following, the potentials were found to be more accurate if all parameters are fitted.

\subsection{Choice of reference data and weights}

\begin{figure}[h!]
  \centering
  \includegraphics[width=0.4 \textwidth]{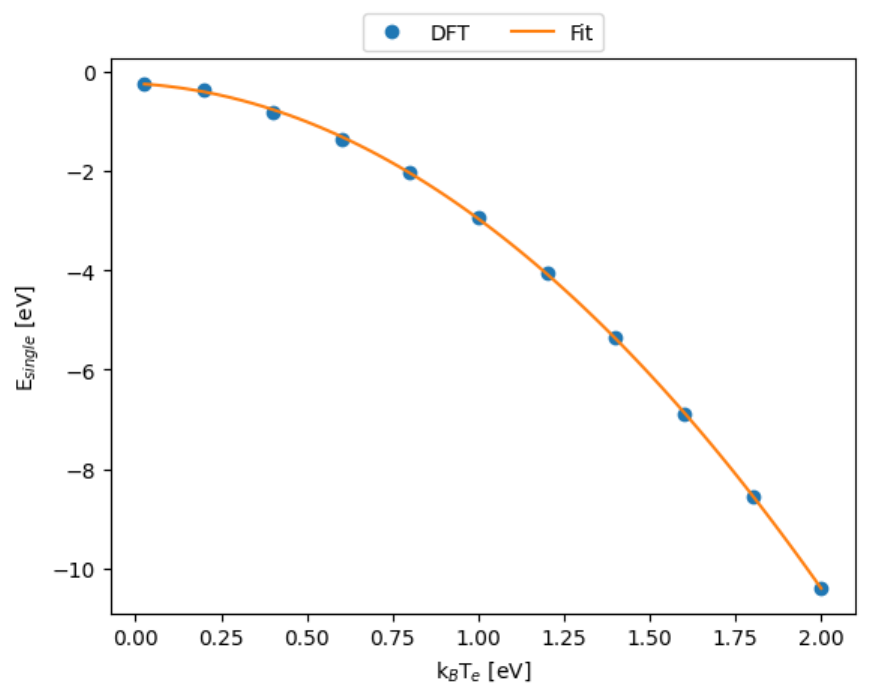}
  \caption{Single-atom energies depending on the electron temperature obtained from DFT calculation and fitted to a second-order polynomial. The fit parameters can be found in \autoref{tab:Fit_parameters}.}
  \label{fig:Single_atom_energies}
\end{figure}

The correct choice of reference data is crucial for any IP. In particular, the reference data has to describe the situation of interest which, in the context of laser ablation, is the material of interest in or close to the ground state structure and during melting at various electron temperatures for the most part. Since there is little to no experimental data of energies, lattice constants or elastic constants at higher electron temperatures that could be used to optimise an IP, DFT data was used as reference data for the IPs. \\
When using potfit for the potential development, one can set weights for different kinds of reference data and also for forces, stresses and energies.
In the case of electron temperature-dependent potentials, cohesive Mermin free energies are used for the fit which are obtained by subtracting the single-atom free energies from the free energies of each respective configuration. These single-atom free energies are shown in \autoref{fig:Single_atom_energies}. \\
The reference data was divided into three groups which, as will be explained later, are weighted differently. All three sets of reference data were used for the potential fitting simultaneously. \\

\begin{itemize}
\item FCC, BCC, SC and HCP structures with all atoms at ideal positions and scaled only in x-direction, scaled in x-, y- and z-direction simultaneously and sheared. 21 of each of these affine transformations on each of the structures were used as reference data. In these transformations, the lattice constant is changed in equidistant steps from $0.8 \cdot a_{0}$ to $1.2 \cdot a_{0}$ in the case of one- or three-dimensional scaling where $a_{0}$ is the relaxed lattice constant of each structure in the unexcited state and the structures are sheared up to \SI{5}{\degree}.
\item A set of FCC reference structures for an accurate description of the elastic constants with smaller deformations and a stricter energy convergence criterion. These structures were created and analysed using the Elastic package \cite{Jochym_1999_Elastic,Jochym_2000_Elastic,Jochym_Elastic_github} with a total of 10 structures scaled by up to \SI{2}{\percent} and sheared up to \SI{2}{\degree}.
\item Non-ideal structures that describe the lattice dynamics at various temperature in the solid state and during melting. For this set, MD simulations were carried out in LAMMPS \cite{Thompson_2022_LAMMPS} using the Embedded Atom Method (EAM) potential developed by Mishin et al. \cite{Mishin_2000_Cu_potential} at 20 temperatures between \SI{100}{K} and \SI{2,000}{K} at zero pressure. After an equilibration phase of 10,000 timesteps of \SI{1}{fs} each, the temperature and pressure were again kept constant and over the next 100,000 timesteps, four snapshots were saved to avoid any correlations between the data. After removing any discrepancies between lattice constants obtained from DFT and MD by rescaling each structure, the respective structures were adopted for a static DFT calculation. Of the four snapshots per temperature, only three were employed for the fitting. The fourth structure will later be used to evaluate the predictive power of the potentials.
\end{itemize}

For the copper potentials to be developed, the weights for all non-FCC structures were set to 1, the weights for the FCC structures with affine transformation to 10, the weights from the calculation of the elastic constants to 100 and the ones from the MD simulations at different temperatures to 10. \\
In total, there are 314 reference structures with a total of 314 energies, 1884 stresses and 77031 forces per electron temperature. There would be a great imbalance in the accuracy of energies, stresses and forces by setting the respective weights of these properties to 1. Therefore, the weights of these three properties are chosen in such a way that they all have approximately the same weight in the end. \\
For each reference structure needed for the potential development, a high cutoff energy of $\SI{700}{eV}$ and a 6x6x6 k-space grid were chosen to include high-energy states and have an accurate sampling of reciprocal space. Furthermore, a total of 900 bands and a convergence criterion of $\SI{1.0e-6}{eV}$ for the structures for the elastic constants and $SI{1.0e-4}{eV}$ for all other structures were used. The best agreement with experimental elastic constants and melting temperatures was found employing a Perdew-Burke-Ernzerhof (PBE) pseudopotential \cite{Perdew_1996_PBE} for copper.
For the comparison to silicon, an Local Density Approximation (LDA) pseudopotential was used as was done in Kiselev's works \cite{Kiselev_2017_Dissertation,Kiselev_2016_HLRS}, 300 to 400 bands depending on the kind of calculation. All other parameters were the same as for copper. \\

\subsection{EAM potential for copper}

An EAM potential \cite{Daw_1983_EAM} is generally described as the sum of a pair potential $\phi \left( r_{ij} \right)$ and an embedding function $F(\rho)$. The latter depends on the transfer function $\rho \left( r_{ij} \right)$ which itself depends on the position of the atoms $r_{ij}$
\begin{equation}
  V = \sum_{i \neq j} \phi \left( r_{ij} \right) + \sum_{i} F_{i} \left[ \rho \left( r_{ij} \right) \right] .
\end{equation}

In the EAM potential for this work, the pair potential is modelled by a Morse potential \cite{Morse_1929_Morse_potential}
\begin{equation}
  \phi(r) = D_{r} \left\{ \left[ 1 - e^{-a \left( r - r_{e} \right)} \right]^{2} - 1 \right\} \Psi \left( \frac{r - r_{\text{cut}}}{h} \right).
\end{equation}
The embedding function is modelled by the universal equation of state proposed by Banerjea et al. \cite{Banerjea_1988_BJS_potential}
\begin{equation}
  F(\rho) = F_{0} \left( 1 - \gamma \ln \rho \right) \rho^{\gamma} + F_{1} \rho .
\end{equation}
Finally, the transfer function is modelled by the function proposed by Chantasiriwan et al. \cite{Chantasiriwan_1996_CSW_potential}
\begin{equation}
  \rho(r) = \frac{1 + a_{1} \cos(\alpha r) + a_{2} \sin(\alpha r)}{r^{\beta}} \Psi \left( \frac{r - r_{\text{cut}}}{h} \right)
\end{equation}
which is supposed to take the Friedel oscillations into account. As was done in the work by Chantasiriwan et al. \cite{Chantasiriwan_1996_CSW_potential}, $\beta = 3$ was not fixed in order to have an additional degree of freedom for better fitting. \\

Discontinuities at the cutoff distance can lead to unphysical forces when an atom is entering or leaving the cutoff radius of another atom. In order to avoid this, a smooth cutoff function
\begin{equation}
  \Psi(x) = \frac{x^{4}}{1 + x^{4}} \hspace{0.5 cm} \text{, where} \hspace{0.5 cm} x = \frac{r - r_{\text{cut}}}{h}
\end{equation}
is used. The cutoff radius was fixed at $r_{\text{cut}} = \SI{6.5}{\angstrom}$ and the parameter responsible for a smooth cutoff $h$ was fixed at $h = \SI{0.5}{\angstrom}$. \\

\begin{figure}[h!]
  \centering
  \begin{subfigure}{0.32 \textwidth}
    \includegraphics[height=0.19 \textheight]{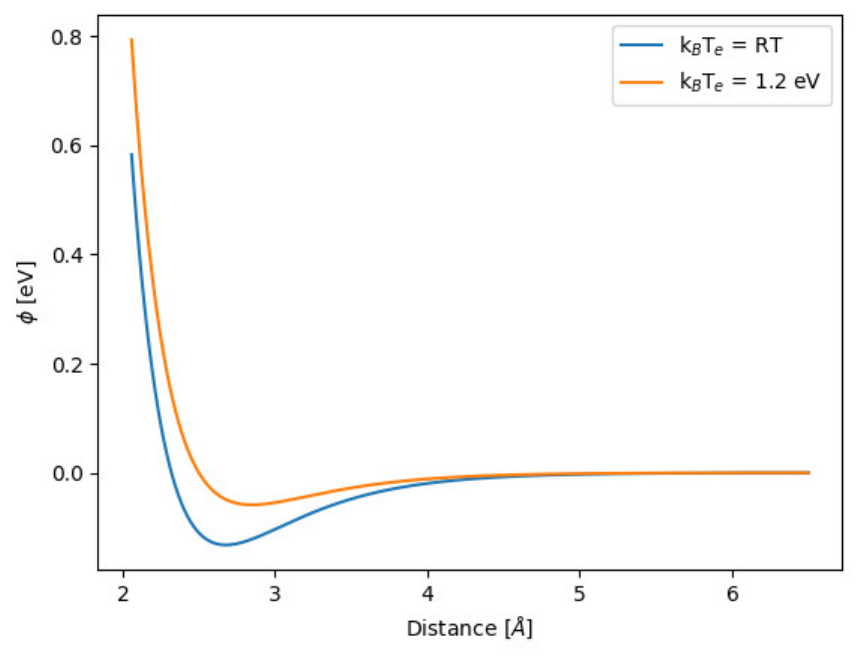}
  \end{subfigure}
  \hfill
  \begin{subfigure}{0.32 \textwidth}
    \includegraphics[height=0.19 \textheight]{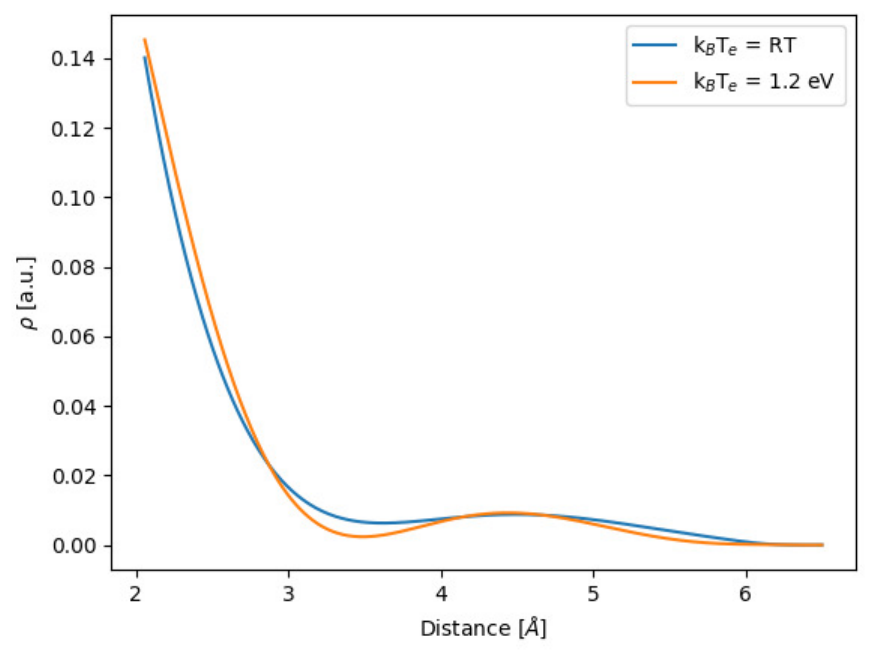}
  \end{subfigure}
  \hfill
  \begin{subfigure}{0.32 \textwidth}
    \includegraphics[height=0.19 \textheight]{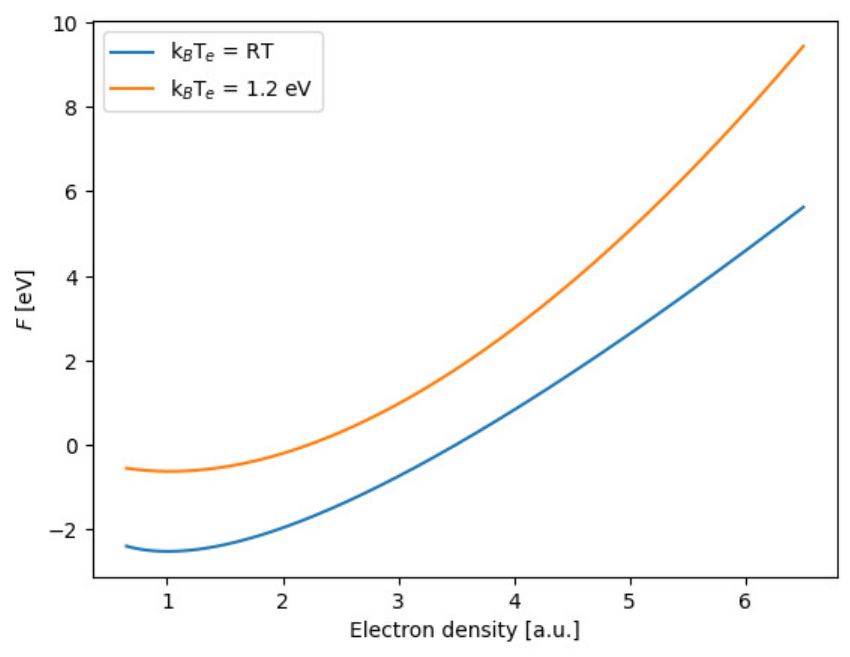}
  \end{subfigure}
  \caption{Pair potential (left), electron density (middle) and embedding function (right) at k$_{B}$T$_{e}$ = RT and k$_{B}$T$_{e}$ = \SI{1.2}{eV}.}
  \label{fig:Potential_functions}
\end{figure}

At higher electron temperatures, the potential fitting becomes increasingly difficult using the method presented here. This is due to the decreasingly prominent minimum of the free energy and finally the complete loss thereof as can already be seen in \autoref{fig:Free_energy}. Still, potentials at \SI{0.025851}{eV} (which from now on, will be indicated by RT), \SI{0.2}{eV}, \SI{0.4}{eV}, \SI{0.6}{eV}, \SI{0.8}{eV}, \SI{1.0}{eV}, \SI{1.2}{eV} were developed independently and with high accuracy. Then, each of the parameters was fitted to a third-order polynomial. The resulting fitted parameters are given in \autoref{tab:Fit_parameters} and shown in \autoref{fig:Fit_parameters} in the appendix (\autoref{sec:Potential_parameters}). \\
The potential functions of unexcited copper and copper at an electron temperature of $\SI{1.2}{eV}$ are shown in \autoref{fig:Potential_functions}. There, as well as in \autoref{fig:Free_energy}, the previously mentioned loss of an energy minimum is displayed. Furthermore, it can already be seen that the potential minimum is shifted to higher values which already hints to an increasing pressure which will be discussed later on.

\begin{table}[h!]
  \centering
  \caption{Fit parameters of the single-atom free energies and all potential functions. All values are rounded to 8 digits after the decimal point.}
  \label{tab:Fit_parameters}
  \begin{tabular}{c|c|c|c|c}
    Parameter & $p_{k, 0}$ [1] & $p_{k, 1}$ [1] & $p_{k, 2}$ [1] & $p_{k, 3}$ [1] \\ \hline
    $E_{\text{single}}$ & -0.23359741 & -0.36458457 & -2.36130430  & -            \\
    $D_{e}$             & 0.13176571  & 0.01754115  & -0.09007948  & 0.02056827   \\
    $a$                 & 1.94150279  & 0.00852749  & -0.04182841  & 0.05499910   \\
    $r_{e}$             & 2.67689887  & -0.01704117 & 0.09799671   & 0.02937452   \\
    $a_{1}$             & -0.41482956 & 0.41420238  & -1.07674855  & 1.17071578   \\
    $a_{2}$             & -0.40874416 & -0.16050673 & -0.32624261  & 0.27606939   \\
    $\alpha$            & 2.05308856  & 0.11030380  & -0.15443881  & 0.30432216   \\
    $\beta$             & 3.32980916  & 0.19323350  & -0.39698869  & 0.27890286   \\
    $F_{0}$             & -2.53847324 & 0.90375312  & 0.50504904   & 0.06982969   \\
    $\gamma$            & 0.79349219  & 0.41031020  & -0.80929640  & 0.67703724   \\
    $F_{1}$             & -0.01210700 & -0.01346310 & 0.02690410   & -0.02088765  \\
  \end{tabular}
\end{table}

\section{Validation of electron temperature-dependent potentials}

In order to evaluate the accuracy of electron temperature-dependent potentials, the analysis is split into two parts. First, properties of the non-excited materials are calculated and compared to experimental values. The focus is put on the ability of reproducing the reference data and data that wasn't used in the fitting process, elastic constants, the melting temperature and phonon spectra. For the calculation of the elastic constants in VASP, the Elastic package \cite{Jochym_1999_Elastic,Jochym_2000_Elastic,Jochym_Elastic_github} was used which was adapted to work with the increased electron pressure in the volume of the non-excited materials. The elastic constants from MD simulations were obtained using LAMMPS \cite{Thompson_2022_LAMMPS}. The phonon spectra were determined from DFT calculations using phonopy \cite{Togo_2023_phonopy_1,Togo_2023_phonopy_2} and using its interface phonoLAMMPS \cite{Carreras_phonolammps_github} for the MD calculations. The melting temperatures were calculated at zero-pressure using calphy \cite{Menon_2021_calphy} which employs thermodynamic integration. \\

For the analysis of the excited state, the elastic constants, the melting temperature and certain phonon modes are considered again but this time while fixing the volume to the volume of the unexcited material to replicate the state shortly after laser irradiation. This way, it is investigated how well the interaction potentials can reproduce non-thermal change of the bond strength obtained from DFT calculations. \\

Since one of the goals of this work is the comparison of highly excited copper with highly excited silicon, the discussion of each property in the excited state will be started with silicon, followed by copper.

\subsection{Validation in unexcited state}

\begin{figure}[h!]
  \centering
  \begin{subfigure}{0.32 \textwidth}
    \includegraphics[width=0.95 \textwidth]{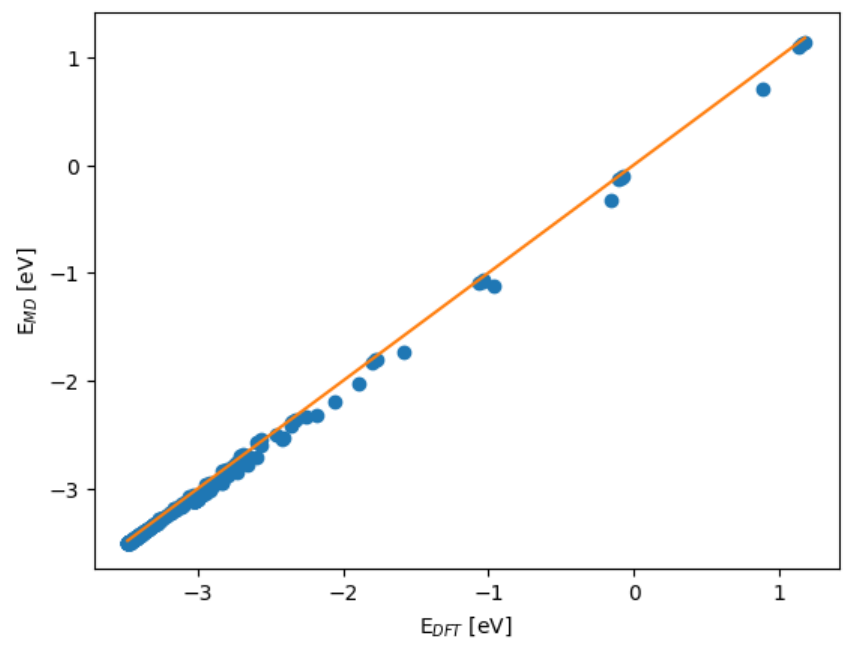}
  \end{subfigure}
  \hfill
  \begin{subfigure}{0.32 \textwidth}
    \includegraphics[width=0.95 \textwidth]{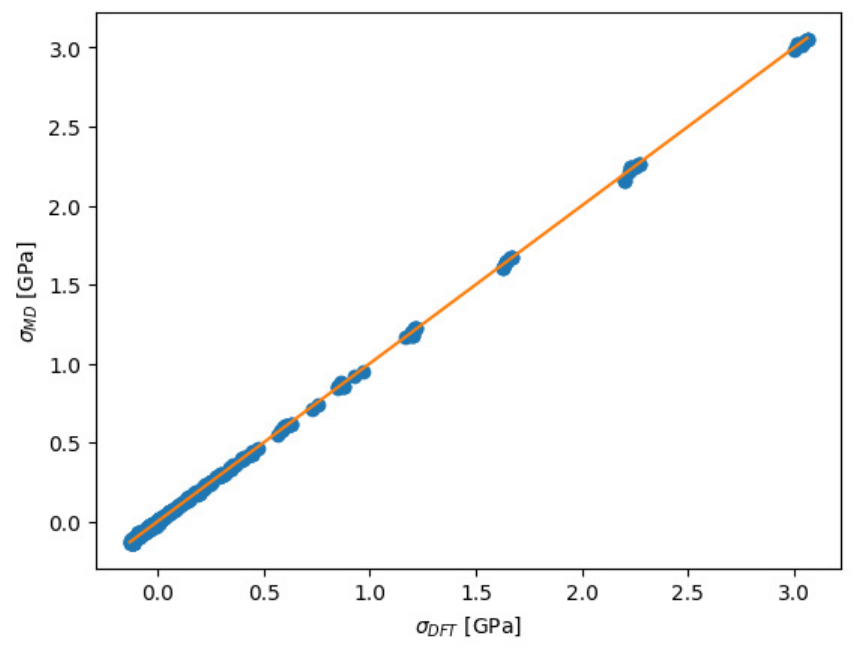}
  \end{subfigure}
  \hfill
  \begin{subfigure}{0.32 \textwidth}
    \includegraphics[width=0.95 \textwidth]{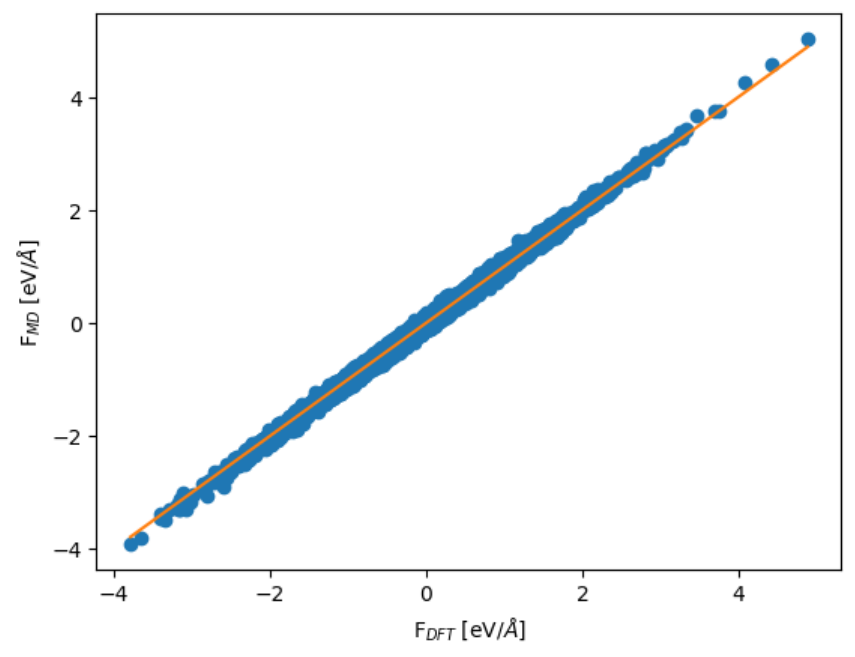}
  \end{subfigure}

  \begin{subfigure}{0.32 \textwidth}
    \includegraphics[width=0.95 \textwidth]{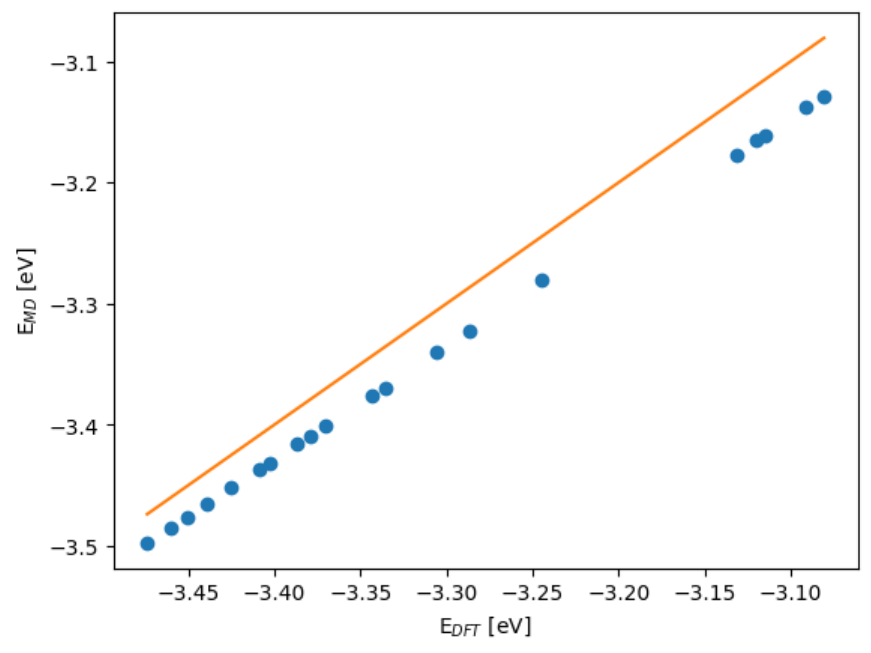}
  \end{subfigure}
  \hfill
  \begin{subfigure}{0.32 \textwidth}
    \includegraphics[width=0.95 \textwidth]{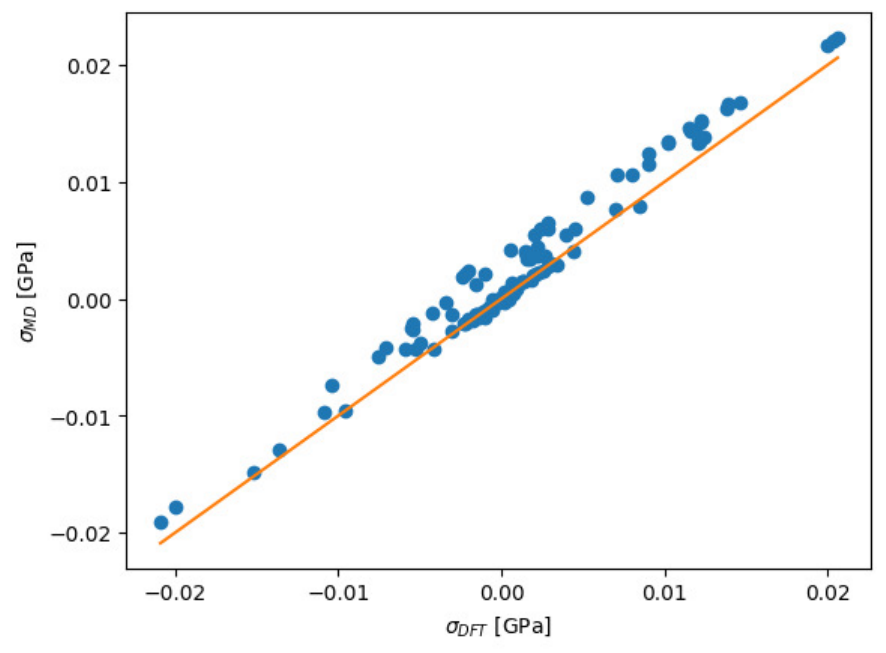}
  \end{subfigure}
  \hfill
  \begin{subfigure}{0.32 \textwidth}
    \includegraphics[width=0.95 \textwidth]{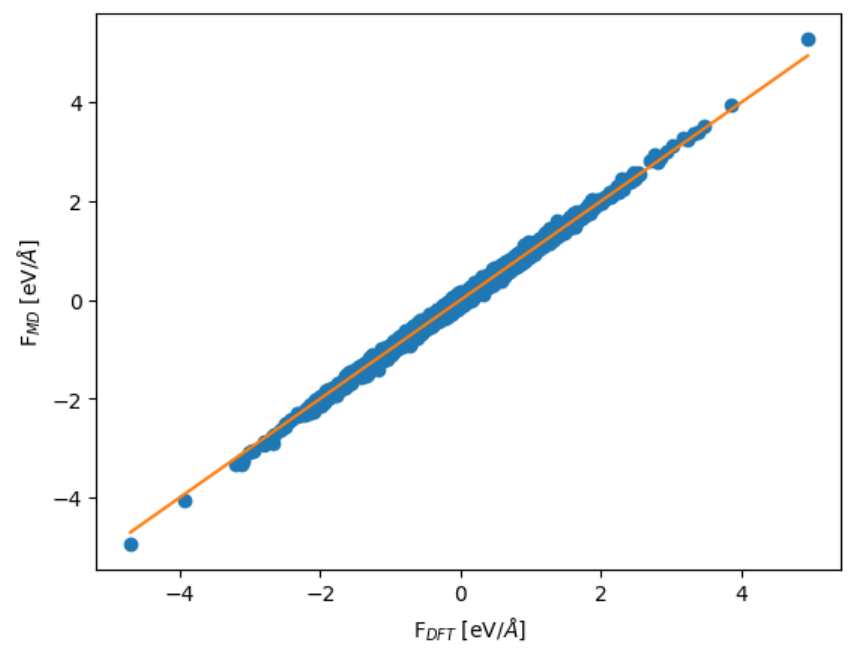}
  \end{subfigure}
  \caption{Comparison of predicted values of energies (left), stresses (middle) and forces (right) using the developed MD potential with DFT data. In the upper plots, the data that was used during the fitting procedure is shown; at the bottom, the predicted values of data that wasn't used during the fit is shown. In each plot, the solid orange line indicates perfect agreement.}
  \label{fig:Comparison_Cu_RT}
\end{figure}

\begin{figure}[h!]
  \centering
  \begin{subfigure}{0.49 \textwidth}
    \includegraphics[width=0.95 \textwidth]{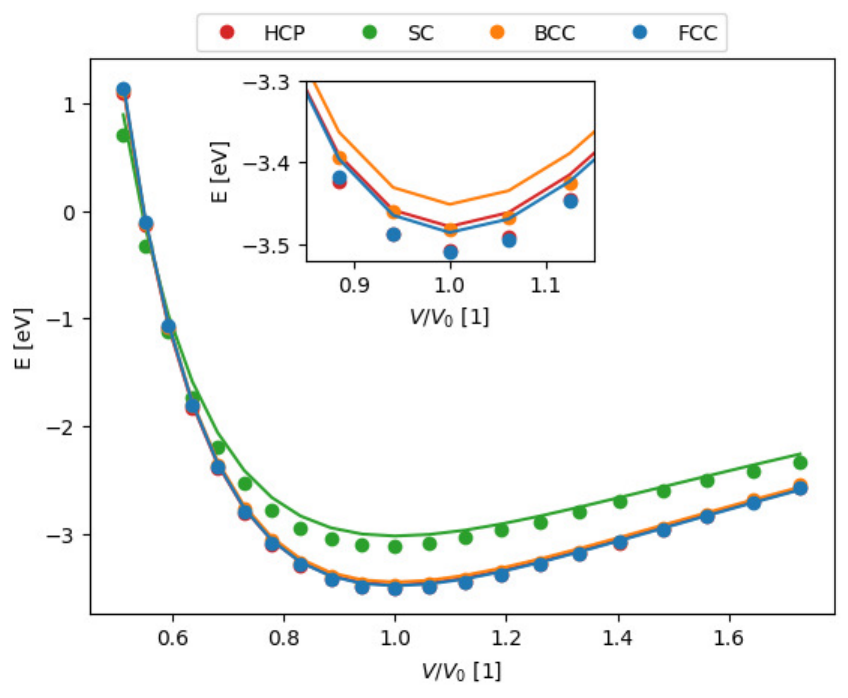}
  \end{subfigure}
  \hfill
  \begin{subfigure}{0.49 \textwidth}
    \includegraphics[width=0.95 \textwidth]{plot_energy_curves_comparison-eps-converted-to.pdf}
  \end{subfigure}
  \caption{Cohesive free energy curves of the HCP, SC, BCC and FCC structure depending on the volume from DFT calculations and predicted by the IP at an electron temperature equivalent to room temperature (left) and at \SI{1.2}{eV} (right). In each plot, the solid lines indicate the DFT data.}
  \label{fig:Comparison_energy_curves_Cu_RT}
\end{figure}

One can investigate the quality of any fitted potential by comparing the predicted values of energies, stresses and forces with the data that was used during the fitting process to show how well it can reproduce energies, stresses and forces. This can be seen in the upper plots of \autoref{fig:Comparison_Cu_RT} for the potential of unexcited. Great accordance between the MD data and DFT data can be found which also remains true for higher electron temperatures. Most plots at higher electron temperatures will be left out here to avoid redundancy. Also, most statements on the quality of the unexcited potential also apply to the excited potentials. \\
Furthermore, in the lower plots of \autoref{fig:Comparison_Cu_RT}, predictive power of the unexcited potential is shown which, again, is very good. There is a consistent underprediction of all the energies which becomes obvious in \autoref{fig:Comparison_Cu_RT} and \autoref{fig:Comparison_energy_curves_Cu_RT}. While optimising the potentials, a particular care was taken to ensure that the FCC structure is the energetically most favourable one since the HCP structure is energetically very close to the FCC structure. And indeed, the order and shape of both cohesive free energy curves are reproduced very accurately over a very large range of volumes.

Next, the elastic constants $C_{11}$, $C_{12}$, $C_{44}$ and the melting temperature of the unexcited materials obtained from DFT calculations and the MD potentials are shown in \autoref{tab:Unexcited_comparison} and compared to experimental values. The bulk modulus is calculated from the elastic constants according to
\begin{equation}
  B = \frac{C_{11} + 2 C_{12}}{3}.
\end{equation}
All elastic constants of copper are within around \SI{10}{\percent} of both, the DFT data and the experimental values, $C_{44}$ deviating a bit more than that. Due to slight underestimation of $C_{11}$ and overestimation of $C_{12}$ compared to the DFT calculations, the bulk modulus can be reproduced almost perfectly. According to the conducted calculations of the melting temperature, the new copper potential underestimates the melting point by around \SI{14}{\percent}. This mismatch will be addressed in more detail when considering the excited state.

\begin{table}[h!]
  \centering
  \caption{Elastic constants $C_{11}$, $C_{12}$, $C_{44}$ and bulk modulus $B$ of silicon and copper from MD simulations with the interaction potentials with degree excitation corresponding to room temperature (\SI{300}{K}) compared to experimental data for copper \cite{Waldorf_1960_elastic_constants_experiment_Cu} measured at \SI{300}{K} and silicon measured at \SI{298}{K}. Furthermore, the melting temperatures $T_{\text{m}}$ are displayed. The experimental melting temperatures of silicon is taken from \cite{Yamaguchi_2002_melting_temperature_Si} and for copper from \cite{Cohen_1966_melting_temperature_Cu}.}
  \label{tab:Unexcited_comparison}
  \begin{tabular}{c|c|c|c|c|c|c}
    Material & Method & $C_{11}$ [GPa] & $C_{12}$ [GPa] & $C_{44}$ [GPa] & $B$ [GPa] & $T_{\text{m}}$ [K] \\ \hline
    \multirow{3}{*}{Si} & DFT & 161.11 & 64.70 & 75.64 & 96.84 & - \\
     & MD & 167.21 & 64.72 & 77.13 & 98.88 & 1673.94 \\
     & Experiment & 165.58 & 63.94 & 79.62 & 97.88 & 1687 \\ \hline
    \multirow{3}{*}{Cu} & DFT & 178.66 & 119.64 & 77.52 & 139.31 & - \\
     & MD & 166.97 & 123.77 & 67.72 & 138.17 & 1168.01 \\
     & Experiment & 168.97 & 120.203 & 75.40 & 136.46 & 1356 \\
  \end{tabular}
\end{table}

\begin{figure}[h!]
  \centering
  \begin{subfigure}{0.49 \textwidth}
    \includegraphics[height=0.28 \textheight]{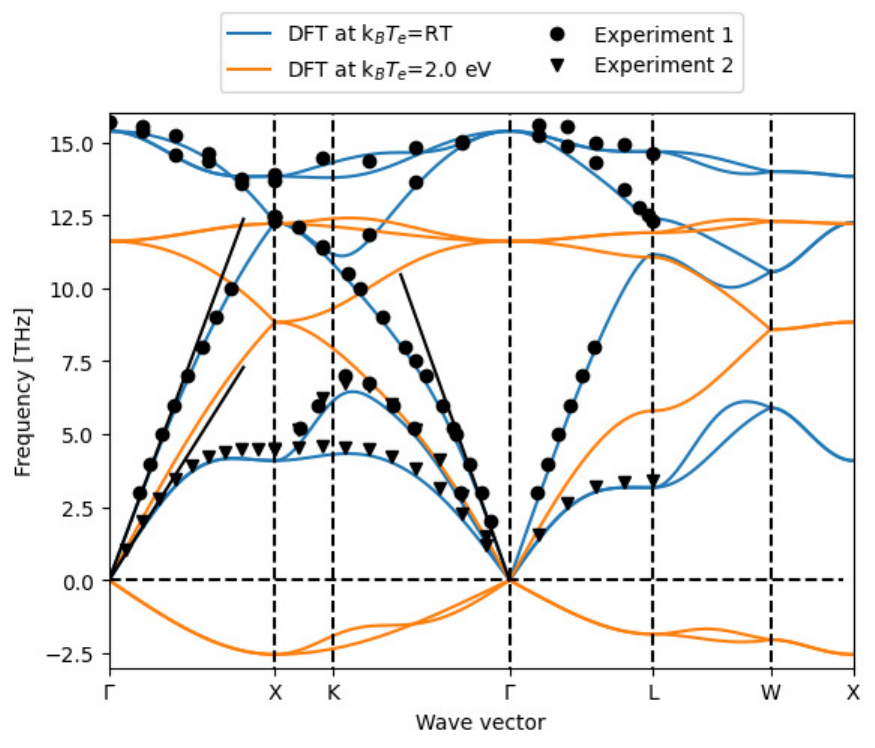}
  \end{subfigure}
  \hfill
  \begin{subfigure}{0.49 \textwidth}
    \includegraphics[height=0.28 \textheight]{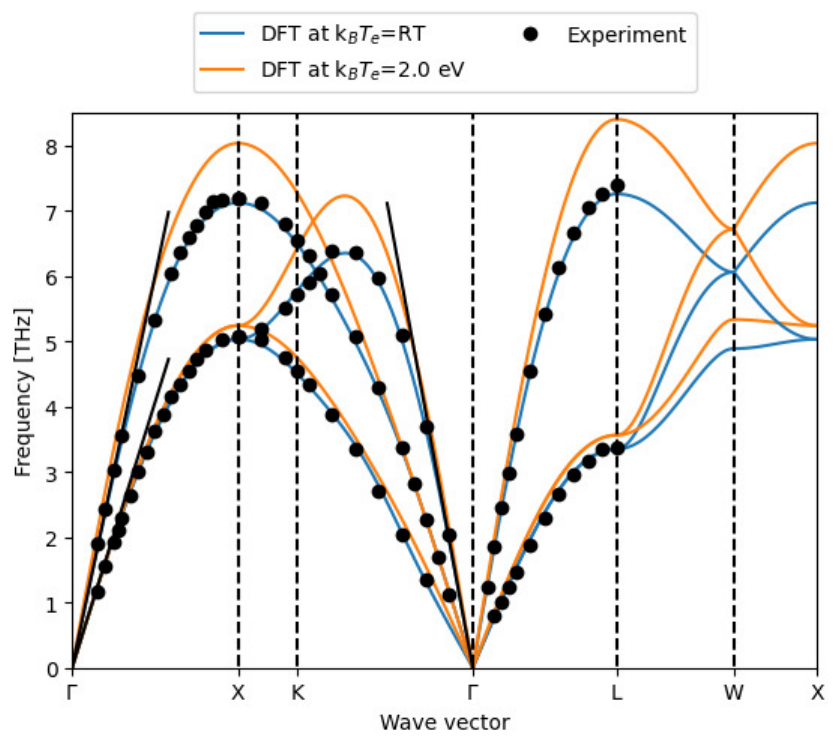}
  \end{subfigure}
  \caption{Phonon spectra at k$_{B}T_{e} =$ RT and k$_{B}T_{e} = \SI{2.0}{eV}$ compared to experimental data. On the left, the comparison between DFT data and results from MD simulations are shown for silicon, on the right for copper. The experimental data for copper was taken from \cite{Svensson_1967_phonon_spectrum_experiment_Cu}, the one for silicon from \cite{Kulda_1994_phonon_spectra_experiment_Si} and \cite{Nilsson_1972_phonon_spectra_experiment_Si}. In both figures, the black lines indicate the longitudinal and transversal sound velocity at k$_{B}T_{e} = RT$.}
  \label{fig:Phonon_spectra}
\end{figure}

Finally and very importantly for the excited material, phonon spectra are compared in \autoref{fig:Phonon_spectra}. In particular, the comparison of the phonon spectra of the unexcited materials with experimental data is shown. Furthermore, the black lines in each plot indicate the longitudinal and transversal speed of sound which will be discussed in more detail for the excited materials. Also, the phonon spectra at a higher degree of excitation is shown in each plot which will also be discussed later on.

\subsection{Validation in excited state}

Here, the effect of the interaction potentials at different degrees of excitation is investigated by looking at possible snapshots following the laser irradiation. This means that in all simulations, the lattice is assumed to have a low temperature and therefore remains in the structure and volume before the excitation. \\

\begin{figure}[h!]
 \centering
 \begin{subfigure}{0.49 \textwidth}
   \includegraphics[height=0.28 \textheight]{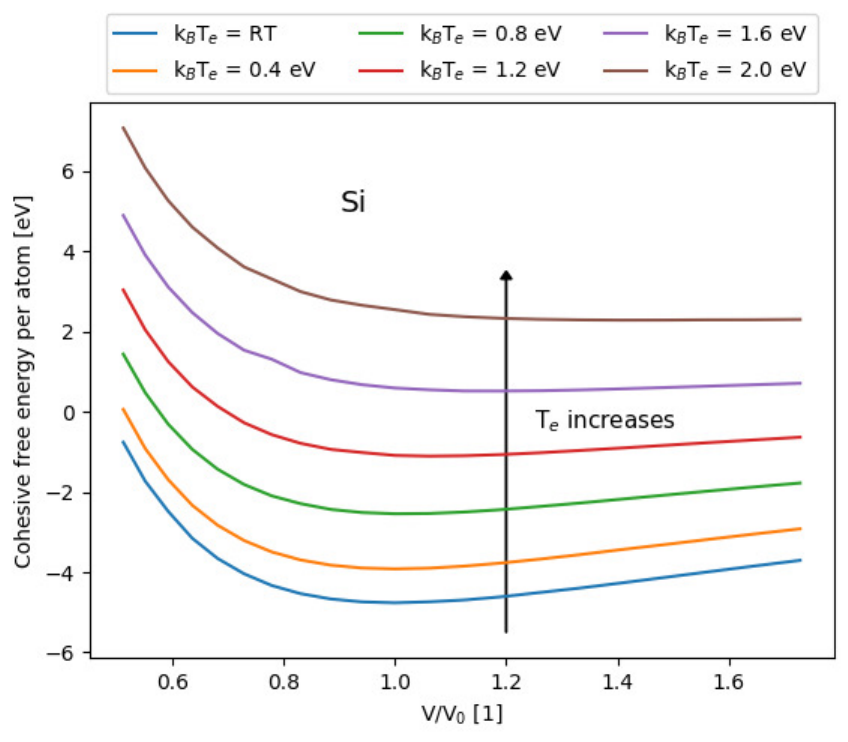}
 \end{subfigure}
  \hfill
  \begin{subfigure}{0.49 \textwidth}
    \includegraphics[height=0.28 \textheight]{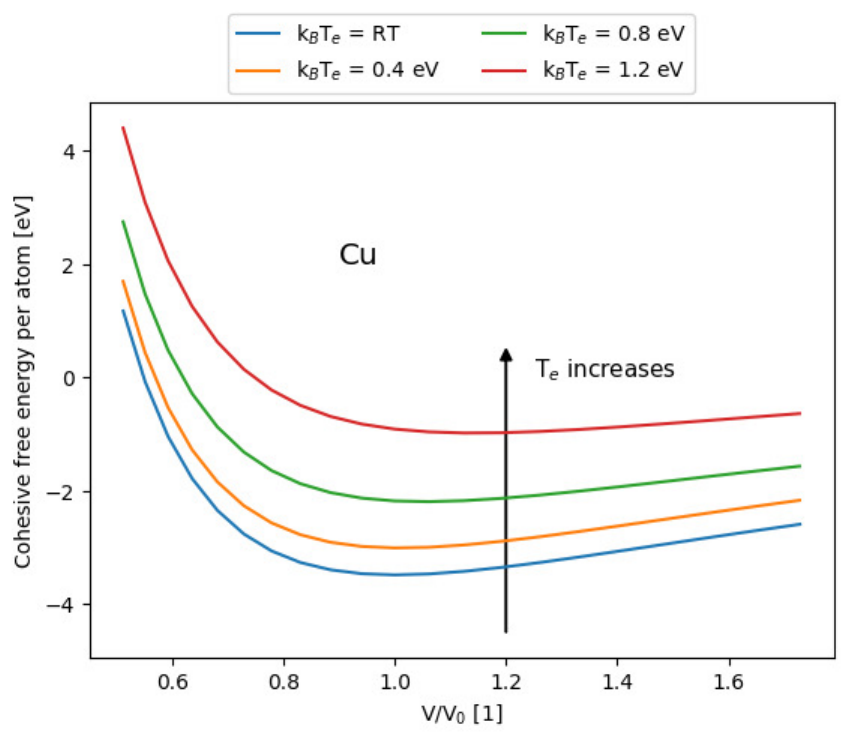}
  \end{subfigure}
  \caption{Cohesive free energy curves of silicon and copper depending on the volume at several different electron temperatures $T_{e}$ obtained from DFT calculations. For silicon, the curve of the diamond structure is depicted and for copper, the curve of the FCC structure is depicted.}
  \label{fig:Free_energy}
\end{figure}

\begin{figure}[h!]
  \centering
  \begin{subfigure}{0.49 \textwidth}
    \includegraphics[height=0.28 \textheight]{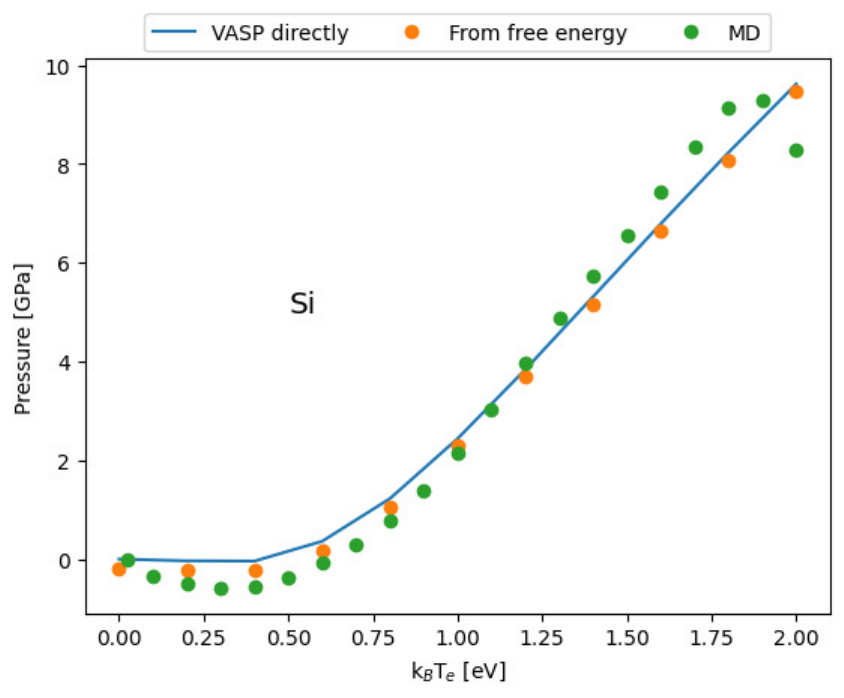}
  \end{subfigure}
  \hfill
  \begin{subfigure}{0.49 \textwidth}
    \includegraphics[height=0.28 \textheight]{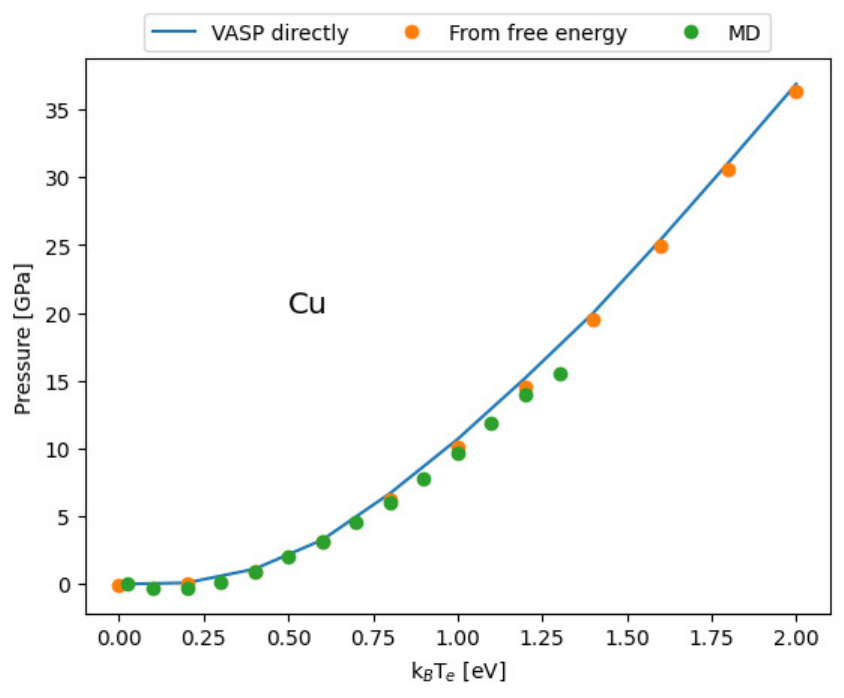}
  \end{subfigure}
  \caption{Pressure of silicon and copper in a simulation box with the volume of the non-excited materials depending on the electron temperature $T_{e}$ obtained from DFT calculations and from MD simulations.}
  \label{fig:Pressure}
\end{figure}

In order to assess the general accuracy of the potentials, the cohesive free energy curves of silicon and copper are shown in \autoref{fig:Free_energy}. For both materials, the cohesive free energy curves are shifted towards lower values due to the higher entropic contribution of the electrons. Furthermore, one can observe the shift of the minimum of each curve towards higher values. This means that the structure will expand with increasing electron temperature if it can. In the context of laser ablation, however, the time scale of strong excitation is short enough that the lattice cannot expand a lot or reorganise. Therefore and as can be seen in \autoref{fig:Pressure}, the pressure near the surface increases. This leads to an increased repulsion of the ablated material as well as a stronger pressure wave through the sample following the expansion near the surface.

Next, in order to quantify the change in bond-strength, the elastic constants and phonon spectra at various electron temperatures are calculated. The elastic constants or rather their change with the degree of excitation contain information about the lattice stability and bond strength according to Born's stability criteria \cite{Born_1940_Lattice_stability}
\begin{equation}
  C_{11} - C_{12} > 0 \hspace{0.5 cm} , \hspace{0.5 cm} C_{11} + 2 C_{12} > 0 \hspace{0.5 cm} \text{and} \hspace{0.5 cm} C_{44} > 0
\end{equation}
or
\begin{equation}
  C_{11} > C_{12} \hspace{0.5 cm}, \hspace{0.5 cm} C_{11} > 0 \hspace{0.5 cm} \text{and} \hspace{0.5 cm} C_{44} > 0.
\end{equation}
This means that an increase of $C_{11}$ and $C_{44}$ indicates an increase of the bond strength, whereas negative values indicate a complete instability of the lattice. Furthermore, similar predictions can be made from the difference of $C_{11}$ and $C_{12}$. \\

\begin{figure}[h!]
  \centering
  \begin{subfigure}{0.49 \textwidth}
    \includegraphics[height=0.28 \textheight]{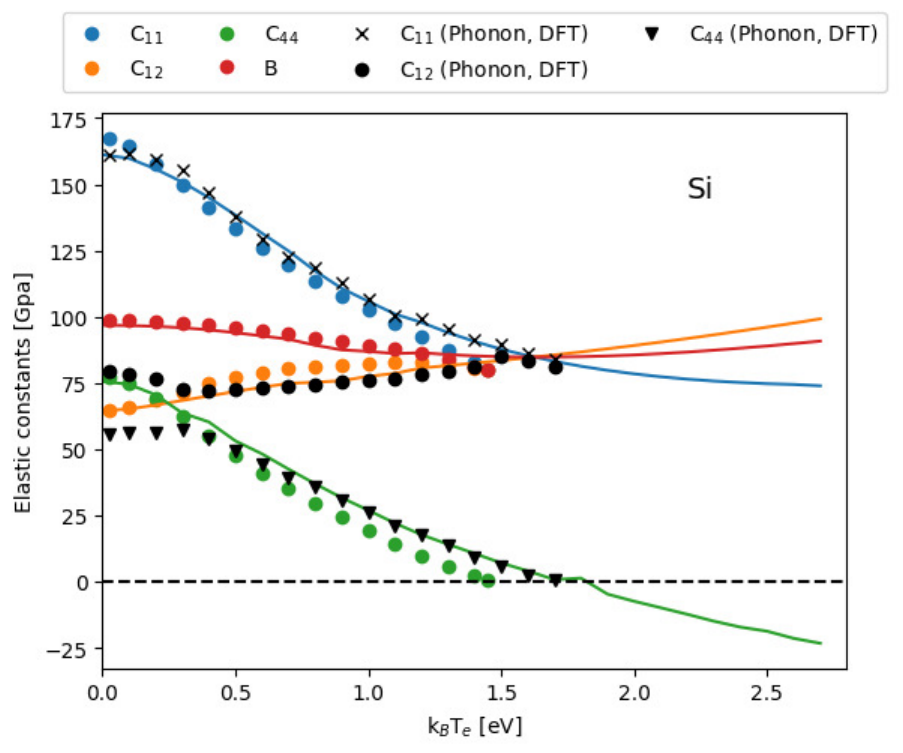}
  \end{subfigure}
  \hfill
  \begin{subfigure}{0.49 \textwidth}
    \includegraphics[height=0.28 \textheight]{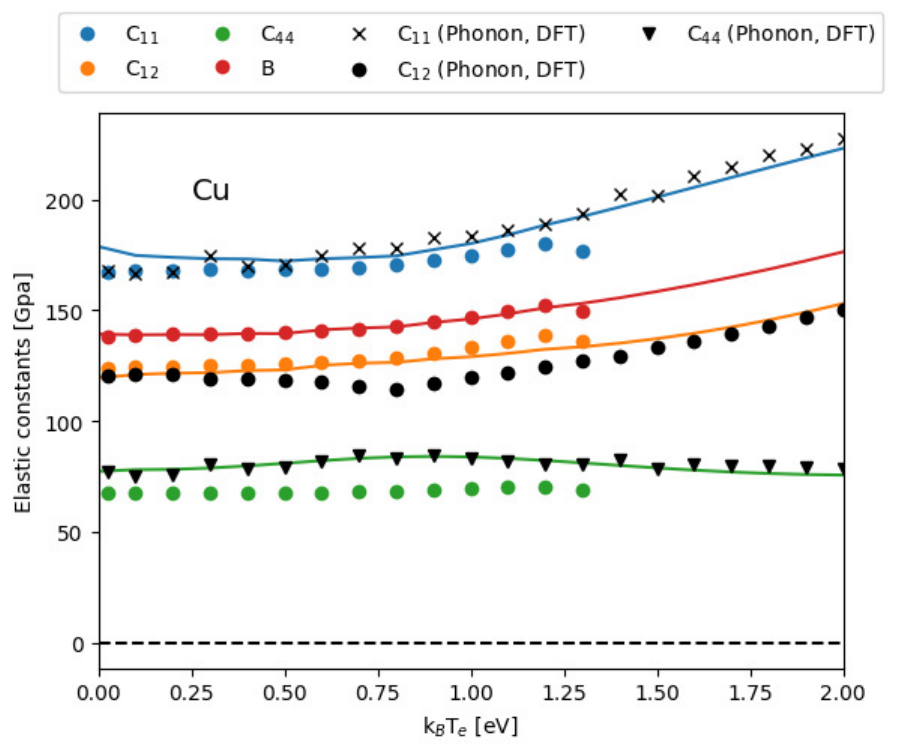}
  \end{subfigure}
  \caption{Elastic constants $C_{11}$, $C_{12}$, $C_{44}$ and bulk modulus $B$ of silicon and copper depending on the electron temperature $T_{e}$. Lines indicate DFT data and dots indicate MD data. In both figures, the black crosses, circles and triangles are calculated elastic constants from the phonon spectra as a proof of consistency.}
  \label{fig:Elastic_constants}
\end{figure}

In \autoref{fig:Elastic_constants}, the elastic constants from DFT and MD calculations are shown. In the case of silicon, the second and third condition for lattice stability are violated above \SI{1.7}{eV} in the case of the DFT calculations and above \SI{1.45}{eV} in the case of the MD calculations. This indicates a lattice instability which leads to the well-known non-thermal melting \cite{Medvedev_2015_nonthermal_Si,van_Vechten_1979_nonthermal_Si}. In the case of copper, however, the opposite can be observed. Both, the elastic constants, indicating bond-hardening.

\begin{figure}[h!]
  \centering
  \begin{subfigure}{0.49 \textwidth}
    \includegraphics[height=0.28 \textheight]{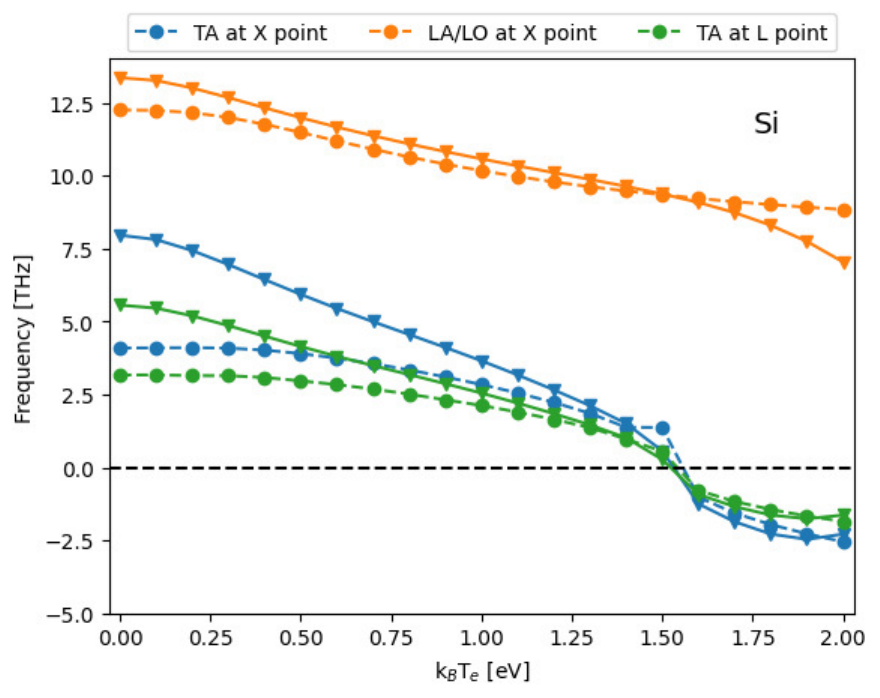}
  \end{subfigure}
  \hfill
  \begin{subfigure}{0.49 \textwidth}
    \includegraphics[height=0.28 \textheight]{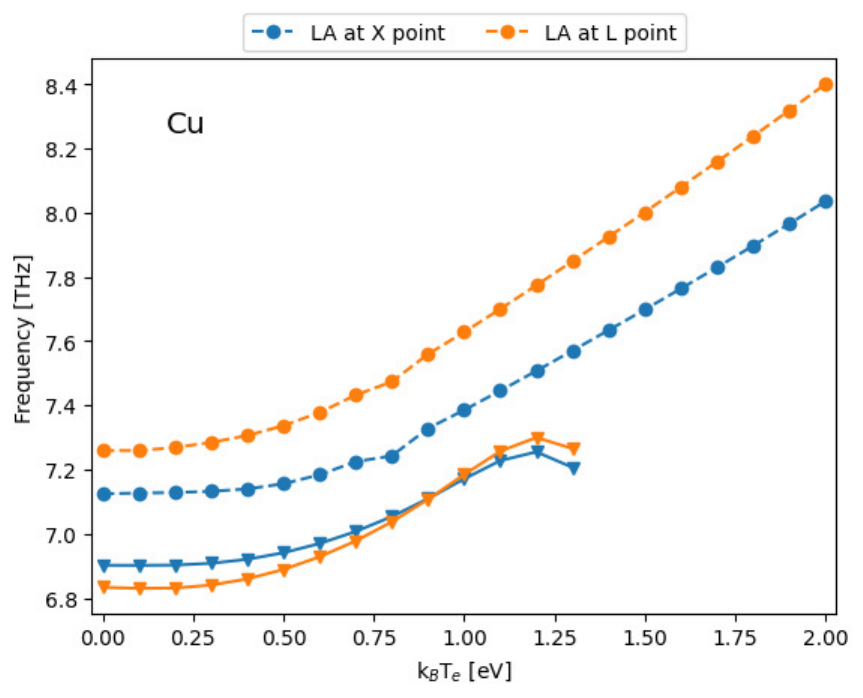}
  \end{subfigure}
  \caption{Phonon frequencies of silicon and copper depending on the electron temperature $T_{e}$ obtained from DFT calculations and MD simulations. For silicon, the TA mode at X (blue) and the TA mode at L (orange) are shown; for copper, the LA mode at X (blue) and the LA mode at L (orange) are shown. Dots indicate DFT data and triangles indicate MD data.}
  \label{fig:Phonon_frequencies}
\end{figure}

To confirm the findings from the elastic constants, the phonon frequencies are calculated next. An increase of phonon frequencies is generally associated with an increase of the bond strength. Vice-versa, negative phonon frequencies hint to an unstable structure. Again, the phonon spectra of the unexcited and the highly excited materials are shown in \autoref{fig:Phonon_spectra}. In \autoref{fig:Phonon_frequencies}, several phonon modes at several high-symmetry points are depicted for silicon and copper over a range of electron temperatures calculated from DFT and MD calculations. While the phonon modes shown in \autoref{fig:Phonon_frequencies} are especially chosen to represent the changes with the electron temperature very clearly, all phonon modes are subject to the same trends as can be seen in \autoref{fig:Phonon_spectra}. Also, for both materials, the MD calculations underestimate the phonon frequencies but the trends with increasing degree of excitation are reproduced very well. \\
In silicon, the TA mode at X and L decrease over the entire range of electron temperatures until it becomes negative at \SI{1.55}{eV} in both, DFT and MD calculations. This is consistent with the elastic constants calculated before. \\
In copper, the phonon modes show a clear increase, again indicating an increase of the bond strength. This is also consistent with the elastic constants.

\begin{figure}[h!]
  \centering
  \begin{subfigure}{0.49 \textwidth}
    \includegraphics[height=0.28 \textheight]{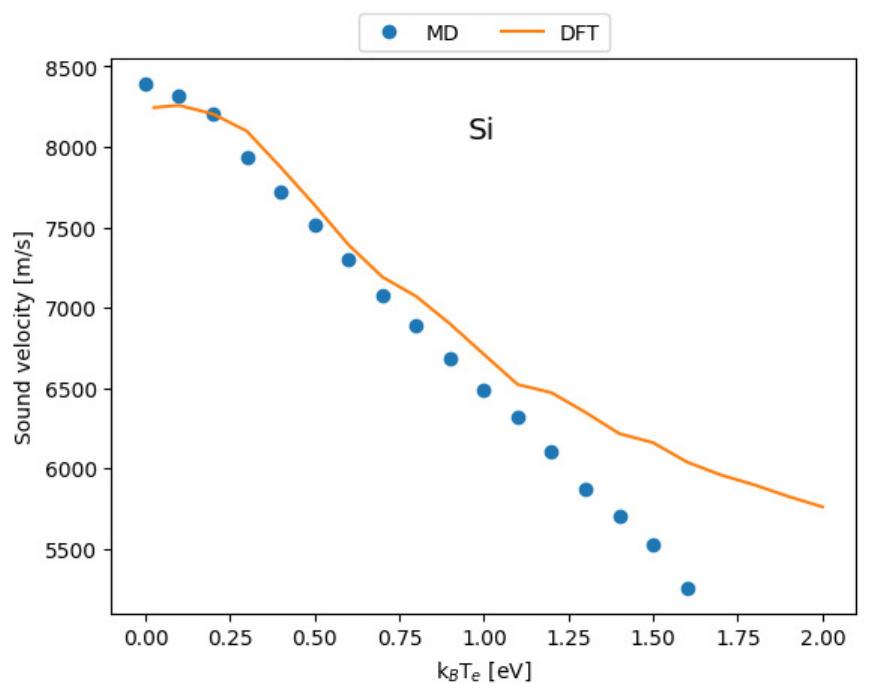}
  \end{subfigure}
  \hfill
  \begin{subfigure}{0.49 \textwidth}
    \includegraphics[height=0.28 \textheight]{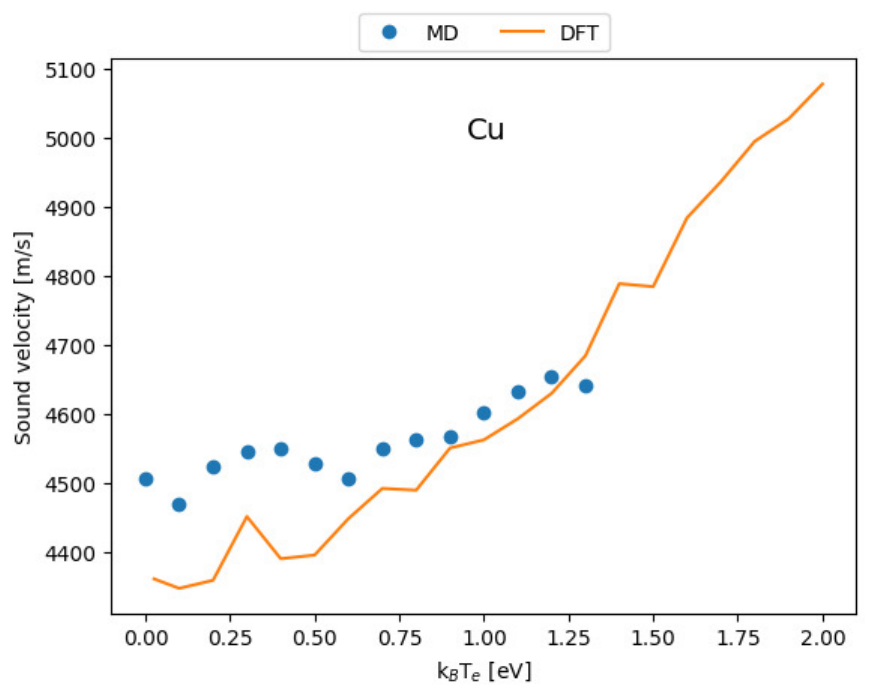}
  \end{subfigure}
  \caption{Sound velocities in of silicon and copper [100] direction depending on the electron temperature obtained from DFT calculations and shock wave MD simulations.}
  \label{fig:Sound_velocities}
\end{figure}

According to \cite{Kittel_2005_Solid_state_physics}, a proof of consistency between the elastic constants and the phonon spectra can be shown via the sound velocities. For that, the connection of the sound velocities in different directions to the elastic constants
\begin{equation}
  \eqalign{v_{\text{s}}^{\text{long}} = v_{\text{s}}^{[100], L} &= \frac{\partial f^{\Gamma \rightarrow X, L}}{\partial k} \at_{\Gamma} = \sqrt{\frac{C_{11}}{\rho}} \cr
  v_{\text{s}}^{\text{trans}} = v_{\text{s}}^{[100], T} &= \frac{\partial f^{\Gamma \rightarrow X, T}}{\partial k} \at_{\Gamma} = \sqrt{\frac{C_{44}}{\rho}} \cr
  v_{\text{s}}^{[110], L} &= \frac{\partial f^{\Gamma \rightarrow K, L}}{\partial k} \at_{\Gamma} = \sqrt{\frac{0.5 \left( C_{11} + C_{12} + 2 C_{44} \right)}{\rho}}}
\end{equation}
is used, where the second upper index of $v_{\text{s}}$ and $f$ indicates the phonon mode considered. The resulting elastic constants are presented in \autoref{fig:Elastic_constants} and the sound velocities in \autoref{fig:Sound_velocities}. The change in the sound velocities has previously been described in \cite{Migdal_2015_FTDFT} for gold and might be relevant for an accurate experimental acoustical monitoring of laser ablation. \\
In order to confirm whether the interaction potentials can reproduce the same, shock wave simulations were conducted using the molecular dynamics code IMD \cite{Stadler_1998_IMD,Roth_2019_IMD} in which the sound velocity was deduced from the propagation of the shock wave front over time. The found sound velocities are compared to the DFT data in \autoref{fig:Sound_velocities}. As can be seen there, the sound velocity in [100]-direction increases slightly with the electron temperature in copper and decreases much more strongly in silicon. In both cases, the MD simulations can reproduce the change of the sound velocities very well. The larger deviation in silicon at higher electron temperatures stems from the structure undergoing melting while the shock wave propagates in the sample which will be discussed in more detail below.

\begin{figure}[h!]
  \centering
  \begin{subfigure}{0.49 \textwidth}
    \includegraphics[height=0.28 \textheight]{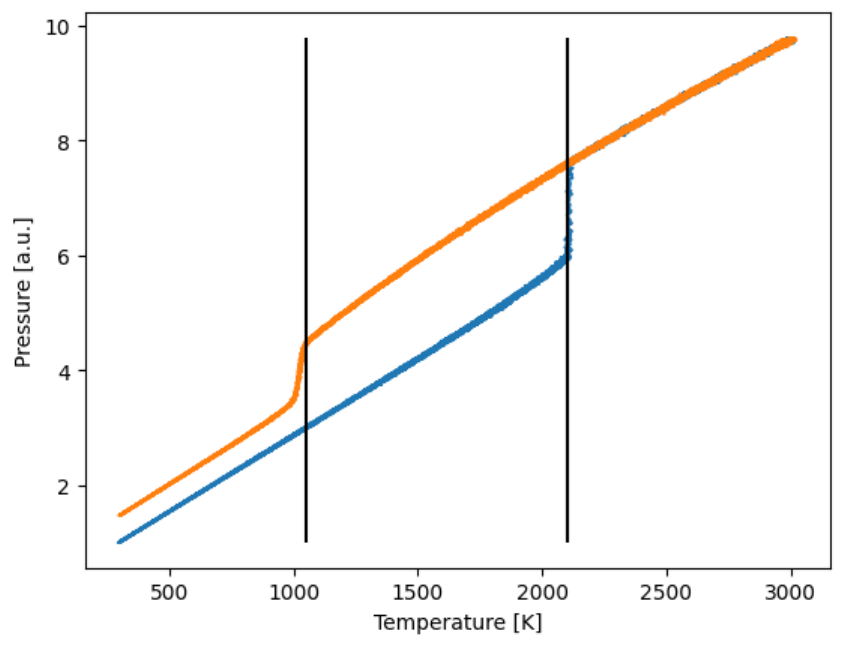}
  \end{subfigure}
  \hfill
  \begin{subfigure}{0.49 \textwidth}
    \includegraphics[height=0.28 \textheight]{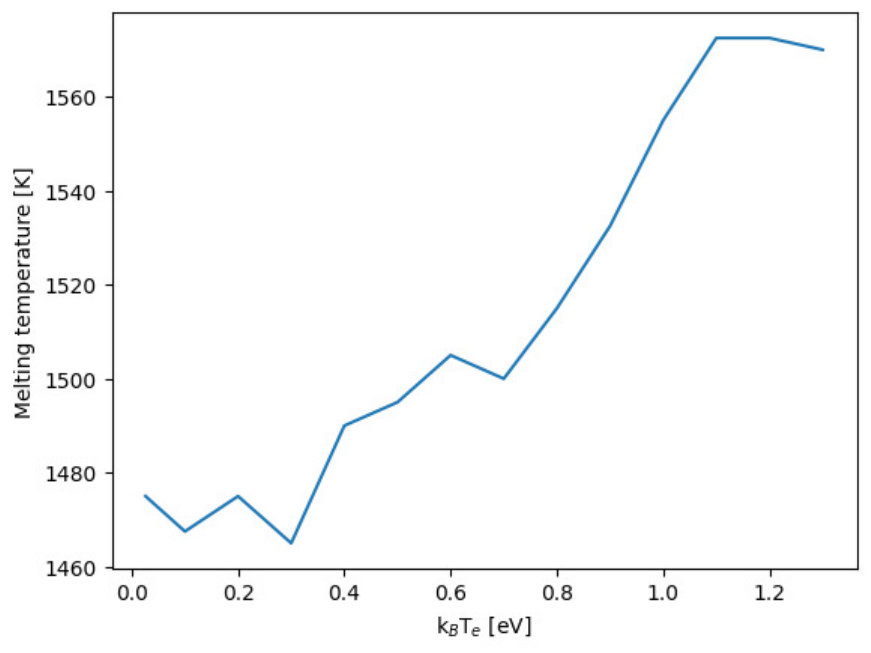}
  \end{subfigure}
  \caption{Left: Pressure change in unexcited copper upon increasing and decreasing the temperature in a simulation box while keeping the volume of the box fixed. The sudden increase of the pressure while heating the box up indicates melting and the sudden decrease of the pressure while cooling the box down indicates solidifying. The starting temperatures of both processes are highlighted by two black lines. Right: Melting temperature of copper depending on the degree of excitation.}
  \label{fig:Melting_temperature_Cu_NVT}
\end{figure}

Finally, the melting temperatures can be determined in a simulation box with fixed volume to replicate the situation right after laser irradiation of the material. To actually obtain the melting temperature, the temperature was first increased and subsequently decreased very slowly, i.e. for five million time steps of \SI{1}{fs}, and the change of the pressure was observed while doing so. A sudden increase of the pressure while increasing the temperature indicates melting whereas a sudden decrease of the pressure while decreasing the temperature indicates the formation of a crystalline structure, therefore solidifying. An exemplary evaluation of the melting temperature of unexcited copper is shown in the left part of \autoref{fig:Melting_temperature_Cu_NVT}. The average of the temperatures at which both processes start is used as a measure of the actual melting temperature. Compared to the results using thermodynamic integration with the respective unexcited potentials, this approach tends to overestimates the melting temperature. \\
For copper, this approach yields an overall increase of the melting temperature with the degree of excitation which is in line with the findings from the elastic constants and the phonon frequencies. In the electron temperature range that can be investigated with the new interaction potential, an increase of the melting temperature of more than \SI{100}{K} can be found compared to unexcited copper. \\

Unfortunately, this can't be applied to silicon since it tends to not recrystallise very well, even with very large samples and very slow cooling. Still, it should be noted that the diamond structure becomes unstable at electron temperatures around \SI{0.8}{eV} electron temperature which is lower than what the change of the elastic constants and the phonon frequencies predict.

\section{Summary and conclusion}

In this work, an interaction potential for copper was developed at various electron temperatures for the use in atomistic Two-Temperature Model simulations to study laser ablation. It was shown that the potential developed in this is capable of reproducing the excitation-induced non-thermal effects, i.e. the bond-hardening and the increasing electron pressure and subsequent expansion very accurately. Explicitly, the change of the elastic constants, the phonon frequencies and the melting temperature predicted by DFT can be reproduced with high accuracy. \\
Since the new potential is based on a previously developed approach that was applied to silicon, it could be confirmed that this approach is highly flexible in terms of materials that can be investigated in this way. While this approach applied to copper turned out to be limited to a maximum electron temperature of around \SI{1.2}{eV}, the potentials showed an impressive accuracy compared to DFT calculations at a fraction of the computational cost. \\
Previously, the inclusion of such non-thermal effects in molecular dynamics simulations has been shown to lead to to a completely different melting dynamics in silicon. In these simulations, the experimental ablation depth and ablation efficiency can only be reproduced accurately when including the change of the interaction potential. Therefore, it remains to be seen how much the increased electron pressure and melting temperature changes the melting dynamics in copper. Furthermore, it will be investigated how the increased pressure compares to inclusion of this pressure via an additional force term as is usually done to incorporate the blast force.

\section{Appendix}

\subsection{Potential parameters}
\label{sec:Potential_parameters}

\begin{figure}[H]
  \centering
  \begin{subfigure}{0.35 \textwidth}
    \includegraphics[width=0.95 \textwidth]{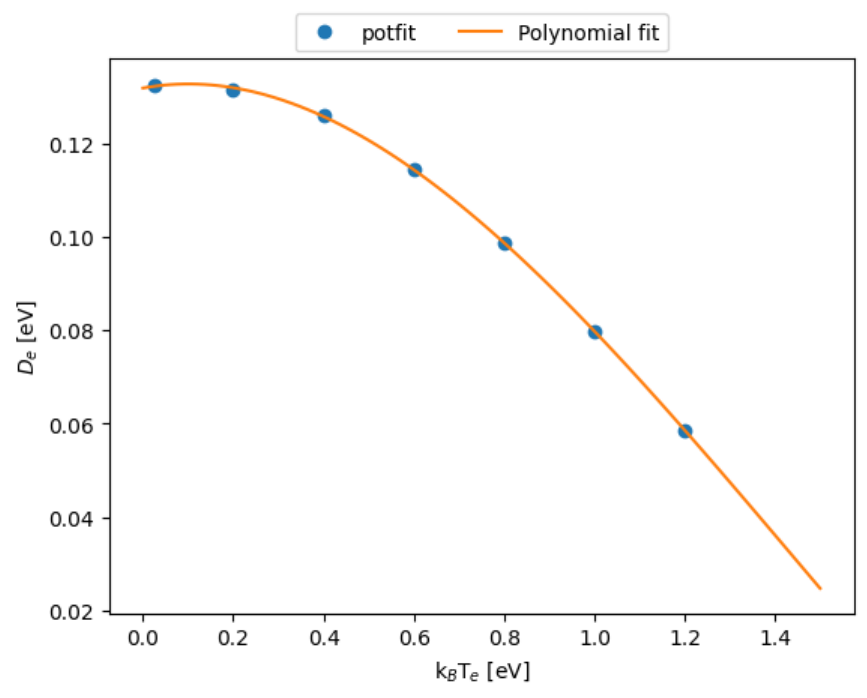}
  \end{subfigure}
  \hfill
  \begin{subfigure}{0.35 \textwidth}
    \includegraphics[width=0.95 \textwidth]{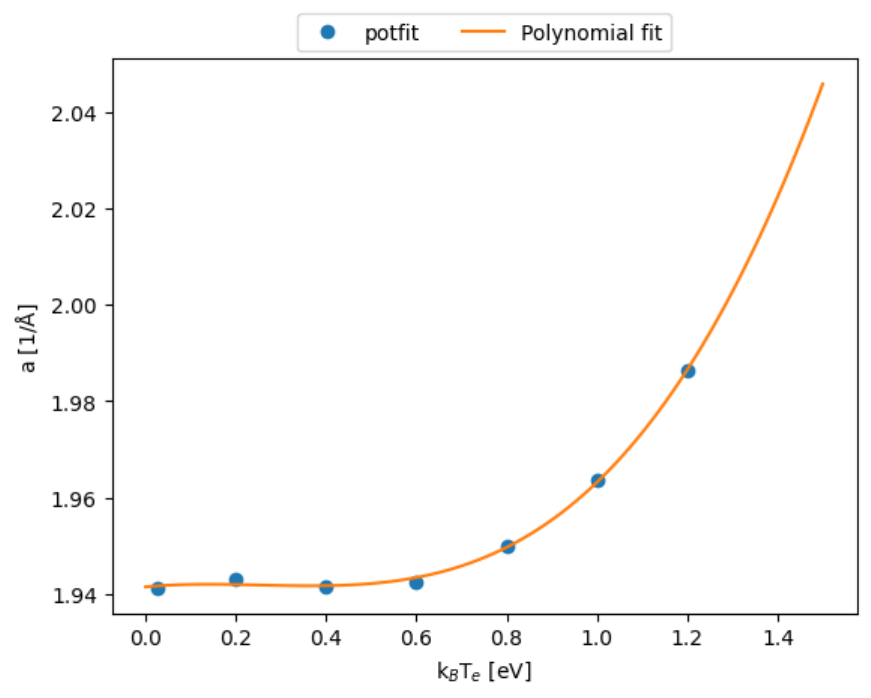}
  \end{subfigure}

  \begin{subfigure}{0.35 \textwidth}
    \includegraphics[width=0.95 \textwidth]{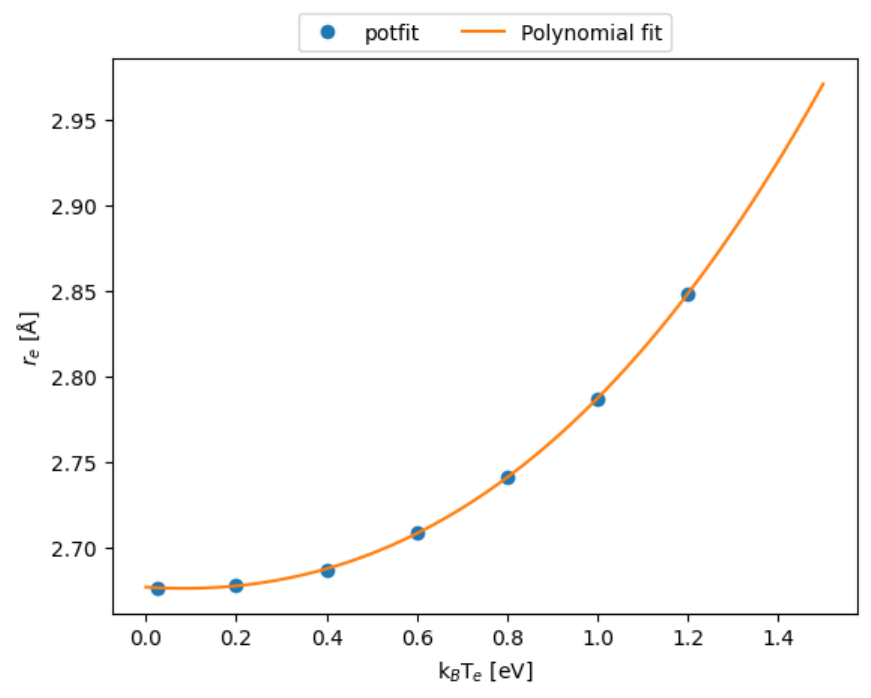}
  \end{subfigure}
  \hfill
  \begin{subfigure}{0.35 \textwidth}
    \includegraphics[width=0.95 \textwidth]{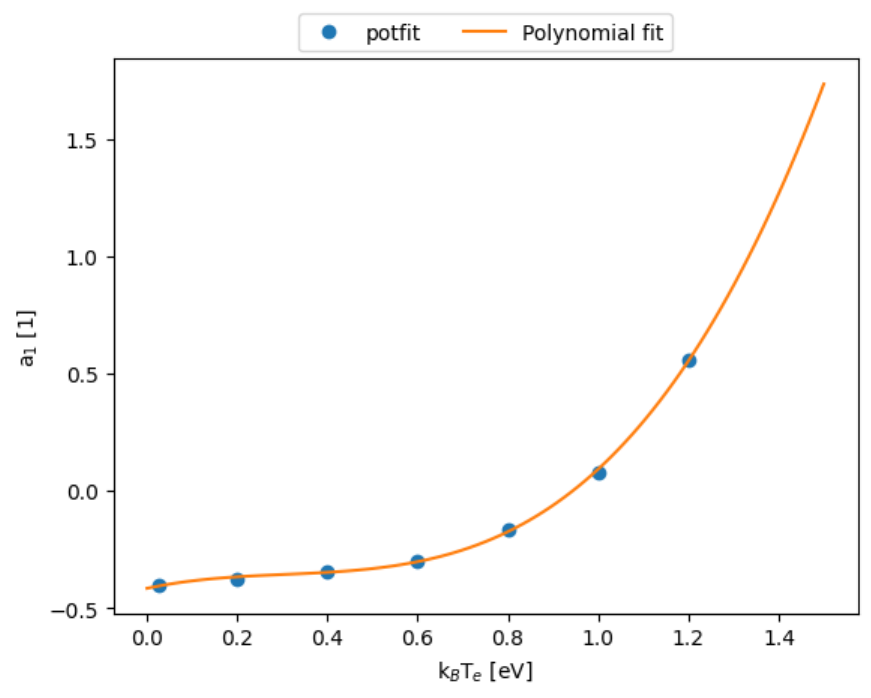}
  \end{subfigure}

  \begin{subfigure}{0.35 \textwidth}
    \includegraphics[width=0.95 \textwidth]{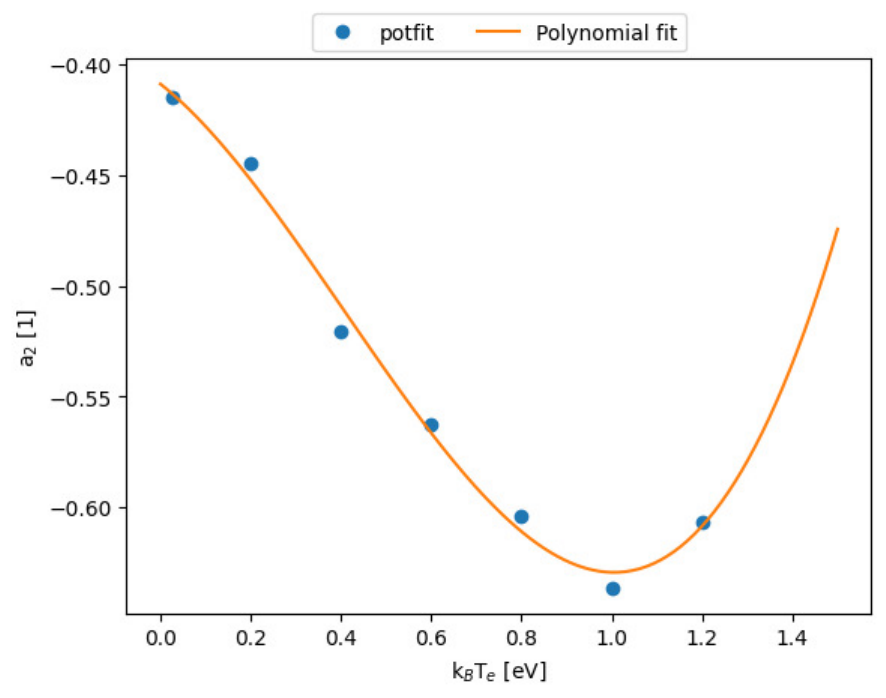}
  \end{subfigure}
  \hfill
  \begin{subfigure}{0.35 \textwidth}
    \includegraphics[width=0.95 \textwidth]{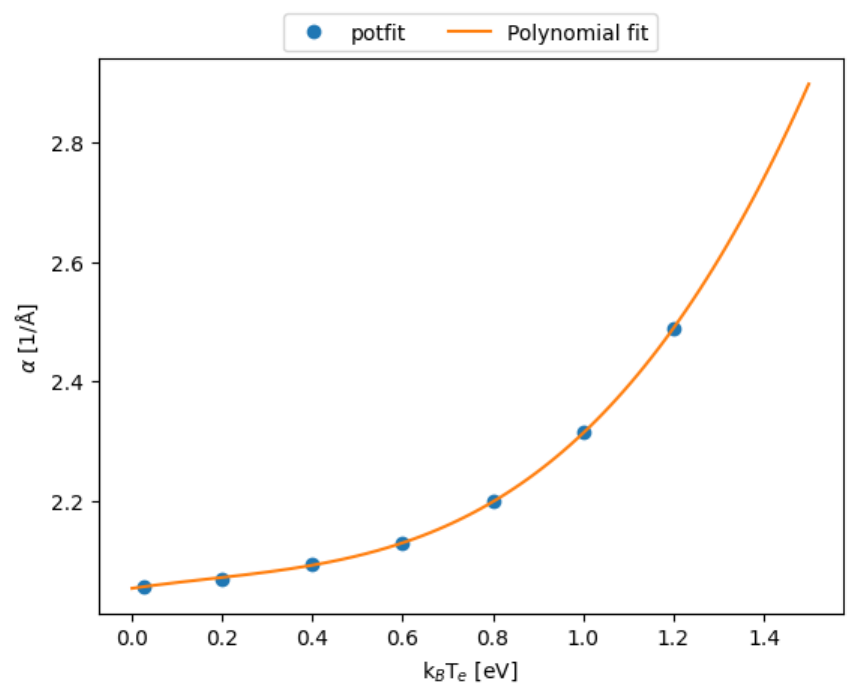}
  \end{subfigure}

  \begin{subfigure}{0.35 \textwidth}
    \includegraphics[width=0.95 \textwidth]{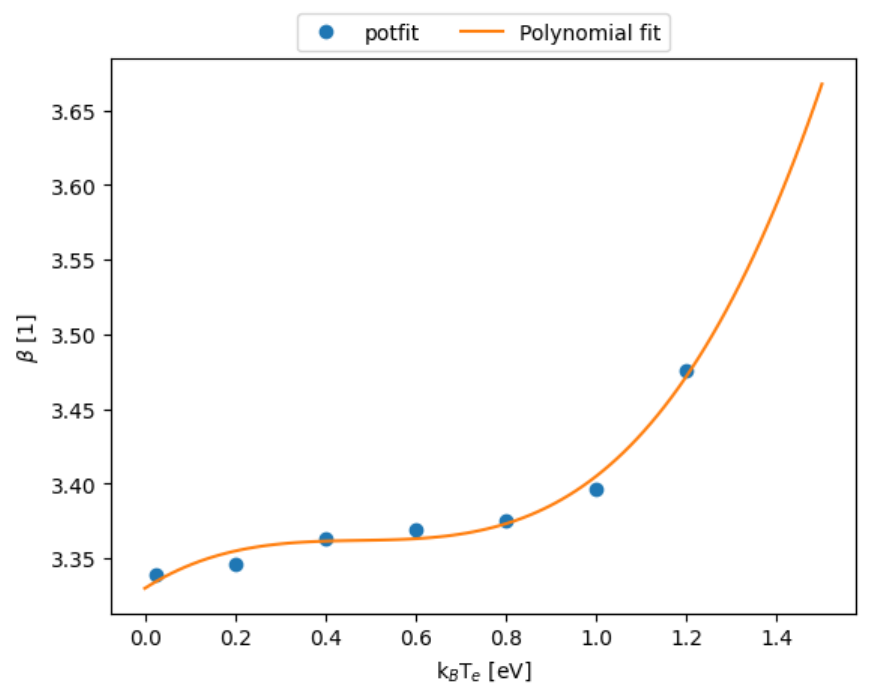}
  \end{subfigure}
  \hfill
  \begin{subfigure}{0.35 \textwidth}
    \includegraphics[width=0.95 \textwidth]{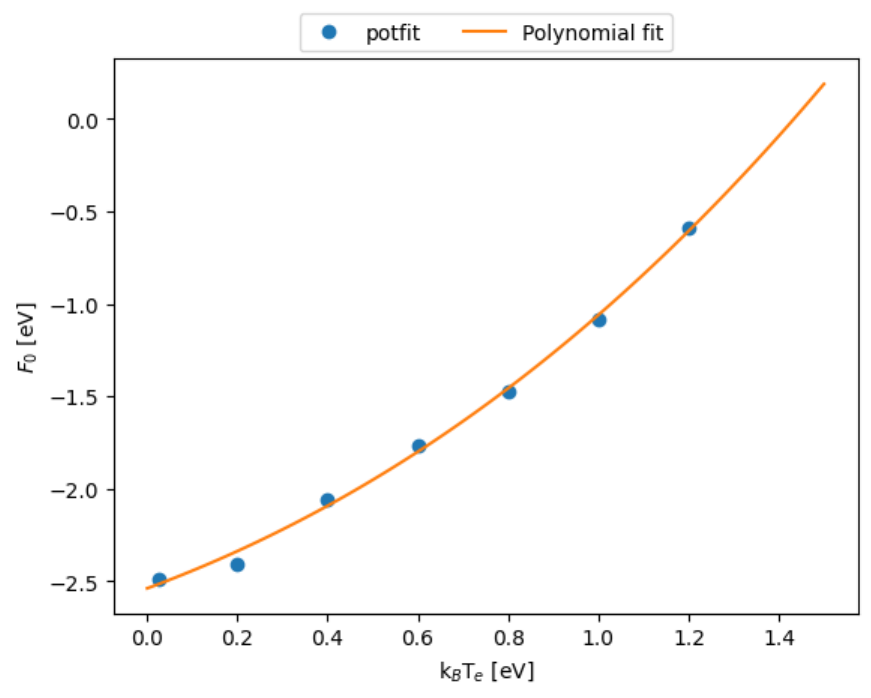}
  \end{subfigure}

  \begin{subfigure}{0.35 \textwidth}
    \includegraphics[width=0.95 \textwidth]{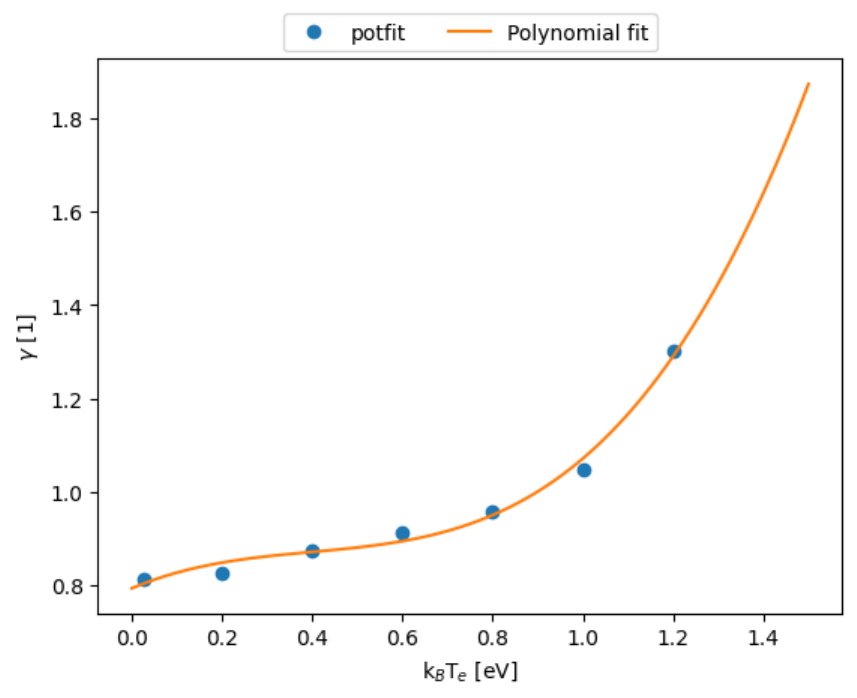}
  \end{subfigure}
  \hfill
  \begin{subfigure}{0.35 \textwidth}
    \includegraphics[width=0.95 \textwidth]{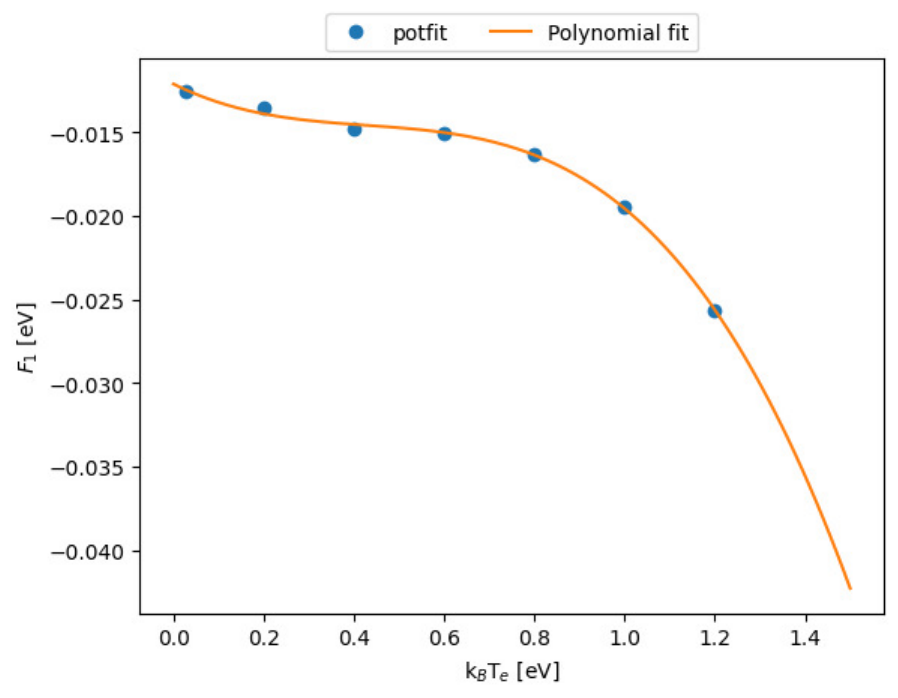}
  \end{subfigure}
  \caption{Potential parameters depending on the electron temperature k$_{B}T_{e}$. The dots represent the fits at each electron temperature, the line represent the interpolation to a fourth-order polynomial. The fit parameters can be found in \autoref{tab:Fit_parameters}.}
  \label{fig:Fit_parameters}
\end{figure}

\subsection{Modified Tersoff potential for silicon}
\label{sec:Tersoff_MOD}

The modified Tersoff potential \cite{Kumagai_2007_Tersoff_mod}

\begin{equation}
  V = \sum_{i \neq j} f_{c} \left( r_{ij} \right) \left[ V_{R} \left(r_{ij} \right) - b_{ij} V_{A}\left(r_{ij} \right) \right]
\end{equation}
with
\begin{equation}
  \eqalign{V_{R} \left(r_{ij} \right) &= A e^{-\lambda_{1} r_{ij}} \hspace{0.5 cm} , \hspace{0.5 cm} V_{A} \left(r_{ij} \right) = B e^{-\lambda_{2} r_{ij}} \cr
  b_ij &= \left( 1 + \zeta_{ij^\eta} \right)^{- \delta} \hspace{0.5 cm} , \hspace{0.5 cm} \zeta_{ij} = \sum_{k \neq i,j} f_{c} \left( r_{ik} \right) g \left( \cos \left( \theta \right) \right) e^{p \left( r_{ij} - r_{ik} \right)^{q}} \cr
  g \left( \cos \left( \theta \right) \right) &= c_{1} + \frac{c_{2} \left( h - \cos \left( \theta \right) \right)^{2}}{c_{3} + \left( h - \cos \left( \theta \right) \right)^{2}} \left[ 1 + c_{4} e^{-c_{5} \left( h - \cos \left( \theta \right) \right)^2} \right]}
\end{equation}

\newpage

and
\begin{eqnarray}
  f_{c} &= 1 \qquad& r_{ij} \le R_{1} \cr
        &= \frac{1}{2} + \frac{9}{16} \cos \left( \pi \frac{r_{ij} - R_{1}}{R_{2} - R_{1}} \right) \cr
        &- \frac{1}{16} \cos \left( 3 \pi \frac{r_{ij} - R_{1}}{R_{2} - R_{1}} \right) \qquad& R_{1} < r_{ij} < R_{2} \cr
        &= 0 \qquad& r_{ij} \ge R_{2}
\end{eqnarray}
was developed by Kiselev et al. \cite{Kiselev_2017_Dissertation,Kiselev_2016_HLRS}.

\section{References}

\bibliographystyle{iopart-num}
\addcontentsline{toc}{chapter}{Bibliography}
\bibliography{references.bib}

\end{document}